\documentclass[10pt,aps,twocolumn,prd,noshowpacs,nofootinbib,noshowkeys,floatfix]{revtex4-1}
\usepackage{graphicx}
\usepackage{bm}
\usepackage{subfigure}
\usepackage{braket}
\usepackage{extarrows}
\usepackage{multirow}
\usepackage{longtable}
\usepackage[normalem]{ulem}
\usepackage{booktabs}
\usepackage[hidelinks]{hyperref}
\usepackage{color}
\usepackage{amssymb,amsmath,amsfonts}
\usepackage{nicefrac}
\usepackage{upgreek}
\usepackage{slashed}
\hypersetup{
    colorlinks,
    linkcolor={blue},
    citecolor={red},
    urlcolor={green}
}

\begin{document}

\title{ Roberge-Weiss transitions at different center symmetry breaking patterns in a $\mathbb{Z}_{3}$-QCD model }
\author{Xiu-Fei Li }
\author{Zhao Zhang}
\email{zhaozhang@pku.org.cn}
\affiliation{School of Mathematics and Physics, North China Electric Power University, Beijing 102206, China}
\date{\today}

\begin{abstract}

We study how the Roberge-Weiss (RW) transition depends on the pattern of center symmetry breaking 
using a $\mathbb{Z}_{3}$-QCD model. We adopt flavor-dependent quark imaginary chemical potentials, 
namely $(\mu_{u},\mu_{d},\mu_{s})/iT=(\theta-2\pi{C}/3,\,\theta,\,\theta+2\pi{C}/3)$ with $C\in[0,1]$. 
The RW periodicity is guaranteed and the center symmetry of $\mathbb{Z}_{3}$-QCD is explicitly broken 
when $C\neq{1}$ or/and quark masses are non-degenerate. For $N_{f}=3$ and $C\neq{1}$, the RW transition 
occurs at $\theta=\theta_{RW}=(2k+1)\pi/3\,(k\in\mathbb{Z})$, which becomes stronger with decrease of 
$C$. When $C={1}$, the $\theta_{RW}$ turns into $2k\pi/3$ for $N_{f}=2+1$, but keeps $(2k+1)\pi/3$ for 
$N_{f}=1+2$; in both cases, the RW transitions get stronger with the mass mismatch. For other $C\neq{0}$ 
cases, the $\theta_{RW}$'s are not integral multiples of $\pi/3$. We find that the RW transition is 
more sensitive to the deviation of $C$ from one compared to the mass non-degeneracy and thus the strength 
of the traditional RW transition with $C=0$ is the strongest. The nature of RW endpoints and its implications
to deconfinement transition are investigated.

\end{abstract}


\maketitle

\section{Introduction}
Exploration of the quantum chromodynamics (QCD) phase diagram at finite temperature and density is one of
the most challenging subjects in particle and nuclear physics. As a first-principle method, the Lattice QCD
(LQCD) simulations yield many meaningful results at vanishing baryon chemical potential
( see~\cite{DElia:2018fjp} and references therein ). However, it is still unavailable at nonzero real chemical
potential region because of the well-known sign problem~\cite{Kogut}. To evade this difficulty, various methods
have been developed~\cite{Z.Fodor1,Z.Fodor2,C. R. Allton,S. Ejiri,Forcrand,Philipsen}. One useful approach is
the analytic continuation from imaginary to real chemical potential~\cite{Forcrand,Philipsen,M. D'Elia}, in
which the fermion determinant is real and thus free from the sign problem.

Introducing an imaginary chemical potential $\mu_{I}=i\theta T$ in QCD corresponds to replacing the fermion
anti-periodic boundary condition (ABC) by the twisted one. In this case, the partition function satisfies
$Z_{QCD}(\theta)=Z_{QCD}(\theta+2\pi/3)$, which is called the Roberge-Weiss (RW) periodicity~\cite{Weiss}.
Since the $\mathbb{Z}_{3}$ symmetry is broken by dynamical quarks, the effective thermal potentials
$\Omega_\phi (\phi=0,\pm2\pi/3)$ of three $\mathbb{Z}_{3}$ sectors have a shift of $2\pi/3$ each other above
some critical temperature $T_{RW}$ and the physical solution is determined by the absolute minimum of the
three ones. This leads to discontinuity of $d\Omega_{\rm{QCD}}(\theta)/d\theta$ at
$\theta=\pi/3\,\,\text{mod}\, \,2\pi/3\,$, which is known as the RW transition~\cite{Weiss}.

The RW transition is a true phase transition for the $\mathbb{Z}_2$ symmetry. LQCD simulations suggest that
the nature of the RW endpoint may depend on quark masses
\cite{FMRW,OPRW,CGFMRW,wumeng,PP_wilson,wumeng2,nf2PP,cuteri,Bonati:2016pwz}: For intermediate quark masses,
it is a critical end point (CEP), while for large and small quark masses it is a triple point. The latest LQCQ
calculation provides evidence that the RW endpoint transition remains second order, in the $3$D Ising universality
class, in the explored mass range corresponding to $m_\pi\simeq 100, 70$ and 50 \text{MeV} \cite{Bonati12}. The
RW transition has also been investigated in effective models of QCD
~\cite{Sakai,Sakai:2009vb,Yahiro,Morita,H.K,Sugano}. Due to the analogy between $\theta$ and the Aharonov-Bohm
phase, it is proposed that the RW transition can be considered as a topological phase transition \cite{kashiwa}.

Note that special flavor-twisted boundary conditions (FTBCs) can lead to an unbroken $\mathbb{Z}_{N_c}$ center
symmetry. As shown in~\cite{ZN1,ZN6}, for $N_f$ flavors with a common mass in the fundamental representation,
the $SU(N_c)$ gauge theory with $d\equiv gcd(N_f,N_c)>1$ has a $\mathbb{Z}_d$ color-flavor center symmetry when
imposing the $\mathbb{Z}_{N_f}$ symmetric FTBCs on $S^1$. The $\mathbb{Z}_d$ symmetry arises due to the intertwined
color center transformations and cyclic flavor permutations. The QCD-like theory for $N_c=N_f=3$ under such FTBCs
is termed as $\mathbb{Z}_{3}$-QCD~\cite{ZN1}. In this theory, the Polyakov loop is the true order parameter for center
symmetry even fermions appear. $\mathbb{Z}_{3}$-QCD is an interesting and instructive theory which is useful for
understanding the deconfinement transition of QCD~\cite{ZN1,ZN2,ZN3,ZN4,ZN5,ZN6,Liu:2016yij,ZN7}.

As mentioned, FTBCs on $S^1$ in $\mathbb{Z}_{3}$-QCD can be replaced with the standard fermion ABCs by
introducing $\mu_{f}=i\theta_{f} T$ ( shifted by $i2\pi{}T/3$ ). Correspondingly, the center symmetry
of $\mathbb{Z}_{3}$-QCD can be explicitly broken by mass non-degeneracy of quarks, or no equal $2\pi/3$
difference in $\theta_f$, or both if color and flavor numbers are unchanged. Then, some natural and
interesting questions arise:
whether the $\mathbb{Z}_{3}$ symmetry breaking in such ways can lead to RW transitions? How do these RW transitions
depend on the center symmetry breaking ? What are the differences between these RW transitions and the traditional
ones in QCD with a flavor-independent $\mu_I$? Answering these questions may deepen our understanding on the relationship
between $\mathbb{Z}_{3}$ symmetry, RW transition and deconfinement transition. Actually, one advantage of
$\mathbb{Z}_{3}$-QCD is that we can use it to study how the pattern and degree of $\mathbb{Z}_{3}$ symmetry
breaking can affect the nature of RW and deconfinement transitions from a different perspective.

The main purpose of this work is to study how RW transitions depend on the center symmetry breaking patterns
by using a $\mathbb{Z}_{3}$-QCD model. We employ the three-flavor PNJL model~\cite{Fu:2007xc,Fukushima:2008wg}, which
possesses the so called extended $\mathbb{Z}_{3}$ symmetry and can correctly reproduce the RW periodicity~\cite{Sakai2}.
Without loss of generality, the flavor-dependent imaginary chemical potentials
$(\mu_{u},\mu_{d},\mu_{s})/iT=(\theta-2\pi{C}/3,\theta,\theta+2\pi{C}/3)$ with $0\leq C\leq 1$ are adopted, which
guarantees $Z(\theta)=Z(\theta+2\pi /3)$ \cite{Sugano}. When quark masses are non-degenerate or $C\neq1$, the
center symmetry of $\mathbb{Z}_{3}-QCD$ is explicitly broken and the RW transitions should appear at high temperature.
We focus on five types of center symmetry breaking and study impacts of variations of $C$ and quark masses
on the RW and deconfinement transitions.

The paper is organized as follows. In Sec.~\ref{Sec_2}, $\mathbb{Z}_{3}$-QCD and the PNJL model with flavor-dependent
imaginal chemical potentials are introduced. In Sec.~\ref{Sec_3}, we present the results of the numerical calculation.
Sec.~\ref{Sec_4} gives the discussion and conclusion.

\section{ $\mathbb{Z}_{3}$-QCD and $\mathbb{Z}_{3}$-symmetric PNJL model}
\label{Sec_2}

\subsection{ $\mathbb{Z}_{3}$-QCD}

The $\mathbb{Z}_{3}$ transformation of QCD is defined as
\begin{equation}
q\rightarrow q'=Uq,\quad A_\mu\rightarrow A_\mu{'}=UA_\mu{U^{-1}}+i(\partial_\mu{U})U^{-1},
\end{equation}
where $U$ is an element of the $SU(3)$ group which satisfies the temporal boundary condition
\begin{equation}
U(x_4=\beta,\textbf{x})={z_k}U(x_4=0,\textbf{x}),
\label{z3bd}
\end{equation}
with $z_k=e^{-i2\pi{k}/3}$ being an element of the center group.
Even the QCD partition function $Z_{QCD}$ is invariant under the $\mathbb{Z}_{3}$ transformation,
the original quark ABC
\begin{equation}
q(x_4=\beta,\textbf{x})=-q(x_4=0,\textbf{x})
\end{equation}
is changed into
\begin{equation}
q(x_4=\beta,\textbf{x})=-e^{i2\pi{k}/3}q(x_4=0,\textbf{x}).
\label{modbc}
\end{equation}
Thus the center symmetry is explicitly broken due to~\eqref{modbc}.
This is why the Polyakov-loop is no longer the true order parameter of deconfinement in QCD.

However, the $\mathbb{Z}_{3}$ symmetry can be recovered if one consider the FTBCs~\cite{ZN1}
\begin{equation}
q_f(x_4=\beta,\textbf{x})=-e^{-i\theta_f}q_f(x_4=0,\textbf{x}),
\label{FTBC1}
\end{equation}
with
\begin{equation}
\theta_f=\frac{2\pi}{3}f  \quad (f=-1,0,1),
\end{equation}
instead of the ABC. For convenience, three numbers -1, 0, and 1 are used as the flavor indices.
Under the $\mathbb{Z}_{3}$ transformation, the FTBCs are transformed into
\begin{equation}
q_f(x_4=\beta,\textbf{x})=-e^{-i\theta_f'}q_f(x_4=0,\textbf{x}),
\label{FTBC2}
\end{equation}
with
\begin{equation}
\theta_f'=\frac{2\pi}{3}(f-k)  \quad (f=-1,0,1).
\end{equation}
We can see that the modified boundary conditions ~\eqref{FTBC2} return to the original ones~\eqref{FTBC1}
if the flavor indices $f-k$ are relabeled as $f$. This means the QCD-like theory with the FTBCs~\eqref{FTBC1}
is invariant under the center transformation if three flavors have a common mass. As mentioned, such a theory
is termed as $\mathbb{Z}_{3}$-QCD, which equals to QCD when $T\rightarrow{0}$.

The FTBCs~\eqref{FTBC1} can be changed back into the standard ABCs through the field transformation~\cite{Weiss}
\begin{equation}
{q_f}\rightarrow{e^{-i\theta_f{T}{x_4}}}q_f,
\label{bctomu}
\end{equation}
which gives rise to the flavor-dependent imaginal chemical potentials
\begin{equation}
{\mu_f}=i\theta_f{T}.
\label{imamu}
\end{equation}
This implies that the global $SU_V(3)\otimes{SU_A(3)}$ symmetry in the
chiral limit is broken to $(U_V(1))^2\otimes{(U_A(1))^2}$ in $\mathbb{Z}_{3}$-QCD \cite{ZN4}.

Equation~\eqref{modbc} and the transformation between FTBCs~\eqref{FTBC1} and imaginal chemical
potentials~\eqref{imamu} indicate the RW periodicity: the partition function $Z(\theta_f)$ is
periodic under the shifts
\begin{equation}
\mu_f/i{T}\rightarrow{\mu_f/iT+2\pi/3},
\end{equation}
i.e.
\begin{equation}
Z(\theta_f)=Z(\theta_f+2\pi/3).
\label{RWP}
\end{equation}

\subsection{$\mathbb{Z}_{3}$-symmetry breaking patterns in $\mathbb{Z}_{3}$-QCD}

The center symmetry of $\mathbb{Z}_{3}$-QCD is attributed to three conditions: 1) $N_f=N_c=3$
(namely, $\gcd(N_f,N_c){>1}$); 2) quark masses are degenerate; 3) the dimensionless flavor-dependent
imaginal chemical potentials (normalized by $iT$) form an arithmetic sequence with the common difference
$2\pi/{N_c}$. Correspondingly, the center symmetry of $\mathbb{Z}_{3}$-QCD will be broken explicitly
if anyone of these conditions is not satisfied, which may lead to the RW transition at high temperature.
It is interesting to study how the possible RW transitions depend on the changes of conditions 2 and/or 3
in $\mathbb{Z}_{3}$-QCD by keeping $N_f=N_c=3$.

Here we express the imaginal chemical potential matrix
$\hat{\mu}=\textrm{diag}(\mu_{u},\mu_{d},\mu_{s})={iT}\hat{\theta}$ in terms of two real parameters $\theta$
and $C$, namely
\begin{equation}
\hat{\theta}=\begin{pmatrix}
\theta_{u}&&\\
&\theta_{d}&\\
&&\theta_{s}
\end{pmatrix}
=\begin{pmatrix}
\theta-\frac{2\pi{C}}{3}&&\\
&\theta&\\
&&\theta+\frac{2\pi{C}}{3}
\end{pmatrix},\label{arithmu}
\end{equation}
where $C\in[0,1]$. As mentioned, such a choice of $\hat{\theta}$ ensures the RW periodicity
$Z(\theta)=Z(\theta+2\pi/3)$. We concentrate on the following center symmetry breaking patterns:
(i) $N_f=3$ with varied $C\neq{1}$ (here and after $N_f=3$ means three flavors share the same mass);
(ii) $N_f= 2+1$ (two lighter flavors have the same mass) with $C=1$;
(iii)$N_f= 1+2$ (two heavier flavors have the same mass) with $C=1$;
(iv) $N_f=2+1$ with varied $C\neq{1}$;
(v) $N_f= 1+1+1$ with $C=1$.

For cases (i)-(iii), the thermal dynamical potential $\Omega(\theta)$ is a $\theta$-even function even
the imaginal chemical potentials are flavor-dependent. For the case (i), three flavors are mass-degenerate
and thus
\begin{align}
&\Omega(\theta)=\Omega(\theta_u,\theta_d,\theta_s)\notag\\
&=\Omega(\theta-{2\pi{C}}{/3},\theta,\theta+{2\pi{C}}{/3})\notag\\
&\xrightarrow{\mathcal{C}}\Omega(-\theta+{2\pi{C}}{/3},-\theta,-\theta-{2\pi{C}}{/3})\notag\\
&\xlongequal{u\leftrightarrow s}\Omega(-\theta-{2\pi{C}}{/3},-\theta,-\theta+{2\pi{C}}{/3})\notag\\
&=\Omega(-\theta),\label{evencase1}
\end{align}
where $\xrightarrow{\mathcal{C}}$ stands for the charge conjugation transformation
and $\Omega(\hat{\theta})=\Omega(-\hat{\theta})$ always holds.
For cases (ii) and (iii), two flavors are mass degenerate (e.g., $m_u=m_d$) and thus
\begin{align}
\Omega(\theta)&=\Omega(\theta_u,\theta_d,\theta_s)\notag\\
&=\Omega(\theta-{2\pi}{/3},\theta,\theta+{2\pi}{/3})\notag\\
&\xrightarrow{\mathcal{C}}\Omega(-\theta+{2\pi}{/3},-\theta,-\theta-{2\pi}{/3})\notag\\
&\xlongequal{\theta\rightarrow \theta+{4\pi}{/3}}\Omega(-\theta-{2\pi}{/3},-\theta-{4\pi}{/3},-\theta)\notag\\
&\xlongequal{u\leftrightarrow d}\Omega(-\theta-{4\pi}{/3},-\theta-{2\pi}{/3},-\theta)\notag\\
&\xlongequal{\theta\rightarrow \theta-{2\pi}{/3}}\Omega(-\theta-{2\pi}{/3},-\theta,-\theta+{2\pi}{/3})\notag\\
&=\Omega(-\theta).\label{evencase2}
\end{align}
Note that $\Omega(\theta)=\Omega(-\theta)$ does not hold for cases (iv) and (v).

\subsection{The center symmetric PNJL model}

The Lagrangian of three-flavor PNJL model of QCD in Euclidean spacetime is defined
as~\cite{Fu:2007xc,Fukushima:2008wg}
\begin{align}
 \mathcal{L}=
 &\bar{q}\left(\gamma_{\mu}D_{\mu}+\hat{m}-\hat{\mu}\gamma_{4}\right)q-G_{\rm S}\sum_{a=0}^{8}
 \left[
 (\bar{q}\lambda^a_f q)^2+(\bar{q}i\gamma_5\lambda^a_f q)^2
 \right]
 \notag \\
 &+G_{\rm D}\left[\det_{ij}\bar{q}_i(1+\gamma^5)q_j
 +\det_{ij}\bar{q}_i(1-\gamma^5)q_j\right]
 \notag \\
 &+{\cal U} (\Phi[A],\Phi^{\ast}[A],T),
 \label{Lag_PNJL}
\end{align}
where $D_{\mu}=\partial_{\mu}+ig_{s}\delta_{\mu 4}A^{a}_{\mu}\lambda^a/2$ is the covariant derivative
with the $SU(3)$ gauge coupling $g_{s}$ and Gell-Mann matrices $\lambda^a$, $\hat{m}=\textrm{diag}(m_{u},m_{d},m_{s})$
denotes the current quark mass matrix, and $\hat{\mu}=\textrm{diag}(\mu_{u},\mu_{d},\mu_{s})$ is the quark chemical
potential matrix. $G_{\rm S}$ and $G_{\rm D}$ are the coupling constants of the scalar-type four-quark interaction
and the Kobayashi-Maskawa-'t Hooft determinant interaction~\cite{tHooft,Kobayashi-Maskawa,Kobayashi-Kondo-Maskawa},
respectively.

The Polyakov-loop (PL) potential ${\cal U} \big( \Phi[A], \Phi^*[A], T \big)$ in the Lagrangian~(\ref{Lag_PNJL})
is center symmetric, which is the function of the Polyakov loop $\Phi$ and its conjugate $\Phi^{\ast}$ and $T$.
The quantity $\Phi$ is the true order parameter for center symmetry in pure gauge theory
(and also in $\mathbb{Z}_{3}$-QCD), which is defined as
\begin{equation}
\Phi = \frac{1}{3}{\rm Tr}(L),
\end{equation}
with
\begin{equation}
L( \mathbf{x})=\mathcal{P}\mathrm{exp}\left[i\int_0^{\beta}d \tau
A_4(\mathbf{x},\tau)\right],
\end{equation}
where $\mathcal{P}$ is the path-integral ordering operator. One popular PL potential is the logarithmic one
proposed in~\cite{S.R}, which takes the form
\begin{align}
&{\cal U} \left(\Phi, \Phi^*, T \right)=T^4\left[-\frac{a(T)}{2}\Phi\Phi^{\ast}\right. \notag\\
&\left.+b(T)\ln (1-6\Phi\Phi^{\ast}+4(\Phi^3+\Phi^{\ast 3})-3(\Phi\Phi^{\ast})^2 )\right],
\label{polyakov_potential}
 \end{align}
where
\begin{align}
a(T)=a_{0}+a_{1}\left(\frac{T_{0}}{T}\right)+a_{2}\left(\frac{T_{0}}{T}\right)^2,\ b(T)=b_{3}\left(\frac{T_{0}}{T}\right)^3.
\end{align}
The potential \eqref{polyakov_potential} will be used in our calculation.

In the Polyakov gauge, the matrix $L$ can be represented as a diagonal form in the color space
\begin{equation}
    \begin{split}
 L= \mathrm{e}^{i\beta A_4}={\rm diag}(\mathrm{e}^{i\beta\phi_1},\mathrm{e}^{i\beta\phi_2},\mathrm{e}^{i\beta\phi_3}),
\end{split}
\end{equation}
where $\phi_1+\phi_2+\phi_3=0$.
The mean-field thermodynamic potential of PNJL then reads
\begin{align}
   &\Omega=
   2G_{\rm S}\sum_{f}\sigma^2_{f}
   -4G_{\rm D}\sigma_{u}\sigma_{d}\sigma_{s}
   -\frac{2}{\beta}\sum_{f}\int\frac{d^3\mathbf{p}}{(2\pi)^3}
   \bigl[
   3\beta E_{f}+
   \notag \\
   &
   \ln(1+3\Phi\textrm{e}^{-\beta(E_{f}-\mu_{f})}
   +3\Phi^{\ast}\textrm{e}^{-2\beta(E_{f}-\mu_{f})}
   +\textrm{e}^{-3\beta(E_{f}-\mu_{f})})+
   \notag \\
   &
   \ln(1+3\Phi^{\ast}\textrm{e}^{-\beta(E_{f}+\mu_{f})}
   +3\Phi\textrm{e}^{-2\beta(E_{f}+\mu_{f})}
   +\textrm{e}^{-3\beta(E_{f}+\mu_{f})})
   \bigr]
   \notag\\
   &+\mathcal{U},
   \label{thermo_PNJL}
 \end{align}
 with $\sigma_{f}=\braket{\bar{q}_{f}q_{f}}$ and $E_{f}=\sqrt{\mathbf{p}^2+M^2_{f}}$ $(f=u,d,s)$.
 The dynamical quark masses are defined by
\begin{align}
 M_{f}=m_{f}-4G_{\rm S}\sigma_{f}+2G_{\rm D}\sigma_{f'}\sigma_{f''},
\end{align}
where $f\ne f'$ $ f'\ne f''$,  and $f\ne f''$. As usual, the three-dimensional cutoff $\Lambda$ is
introduced to regularize the vacuum contribution.
For the pure imaginary chemical potential case, we can write $\Phi$ and $\Phi^{*}$ as
\begin{equation}
\begin{split}
\Phi = Re^{i\phi},  \ \ \
\Phi^{\ast} = Re^{-i\phi}.
\end{split}
\label{loop}
\end{equation}
The condensates $\sigma_f$, the magnitude $R$,
and the phase $\phi$ are determined by the stationary conditions
\begin{equation}
\frac{ \partial \Omega}{\partial\sigma_u}
=\frac{ \partial \Omega}{\partial\sigma_d}
=\frac{ \partial \Omega}{\partial\sigma_s}
=\frac{ \partial \Omega}{\partial R}
=\frac{\partial \Omega}{\partial\phi} = 0.
\end{equation}

Similar to $\mathbb{Z}_{3}$-QCD, the three-flavor PNJL with a common quark mass possesses
the exact $\mathbb{Z}_{3}$ symmetry when introducing the special flavor-dependent
imaginary chemical potentials ${\hat{\mu}}={iT}\hat{\theta}$, where
$\hat{\theta}={\rm diag}(\theta-2\pi/3,\theta,\theta+2\pi/3)$~\cite{ZN1}. Here we refer to
this center symmetric PNJL as $\mathbb{Z}_{3}$-PNJL, which can be regarded as a low-energy
effective theory of $\mathbb{Z}_{3}$-QCD. The RW transitions under the conditions (i)-(v)
will be studied in the $\mathbb{Z}_{3}$-PNJL formalism by breaking the center symmetry
explicitly.

\begin{table}[t]
\begin{center}
\caption{The parameter set in the PL potential sector }
\begin{tabular}{cccccc}
\hline
  & $a_{0}$\qquad\qquad   & $a_{1}$\qquad\qquad  & $a_{2}$\qquad\qquad  & $b_{3}$\qquad\qquad & $T_{0}$ [MeV]\qquad \\ \hline
  & 3.51\qquad\qquad      & -2.47\qquad\qquad    & 15.2\qquad\qquad     & -1.75\qquad\qquad   & 195\qquad \\ \hline
\end{tabular}
\label{table1}
\end{center}
 \end{table}

\begin{table}[ht]
\begin{center}
\caption{The parameter set in the NJL sector. }
\begin{tabular}{cccccc}
\hline
  & $m_{u(d)}$ [MeV]  & $m_{s}$ [MeV]     & $\Lambda$ [MeV]     & $G_{\rm S}\Lambda^2$       & $G_{D}\Lambda^5$ \\ \hline
  & 5.5               & 140.7             & 602.3               & 1.835                      & 12.36 \\ \hline
\end{tabular}
\label{table2}
\end{center}
\end{table}

\begin{figure}[htbp]
\centering
 \includegraphics[width=0.9\columnwidth]{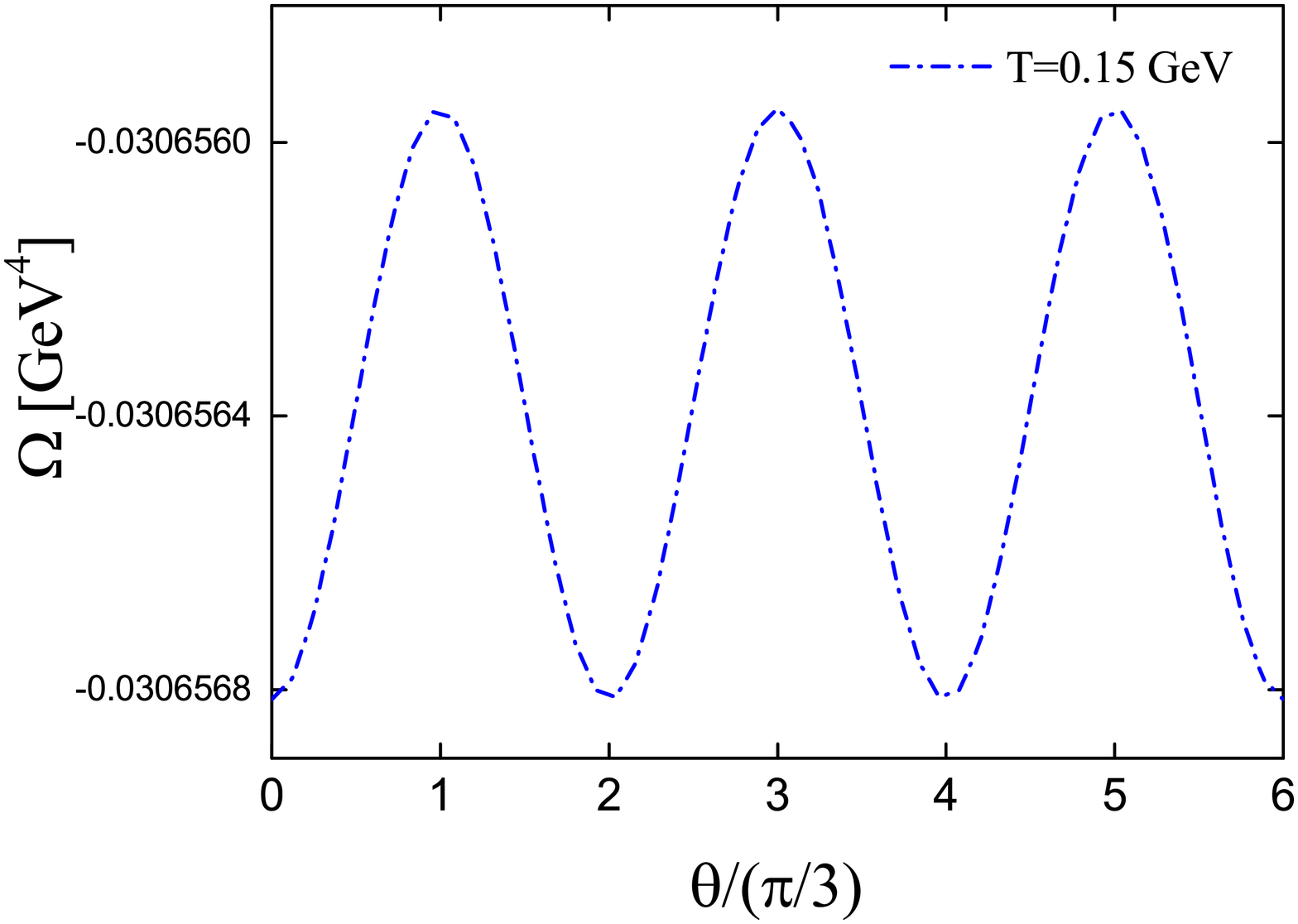}
 \includegraphics[width=0.9\columnwidth]{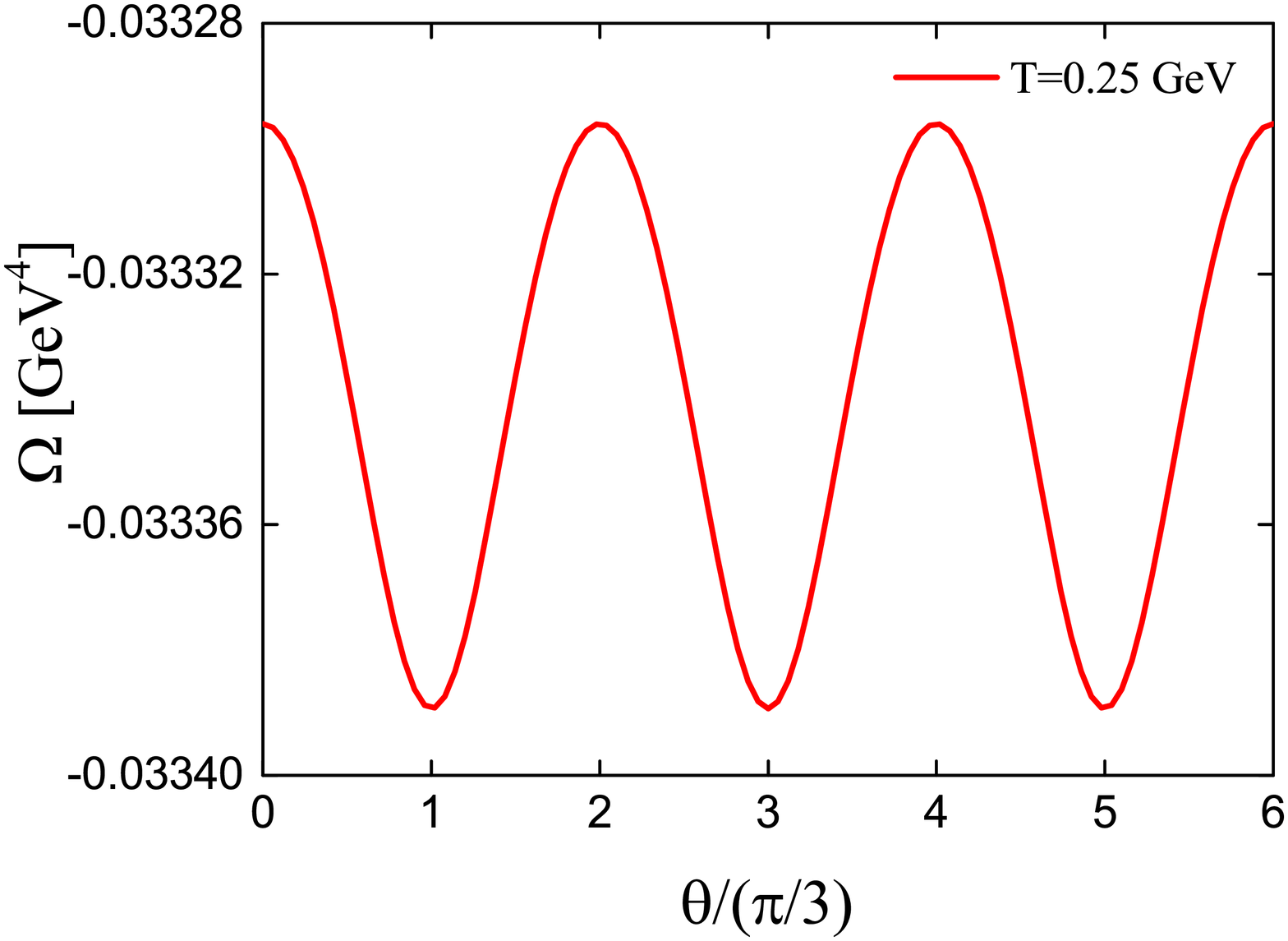}
 \caption{Thermodynamic potential $\Omega$ as the function of $\theta$ for $T=150$ MeV (upper) and $T=250$
 MeV (lower) in $\mathbb{Z}_3$-PNJL with $m_u=m_d=m_s=5.5$ MeV and $C=1$.}
 \label{fig:1}
\end{figure}

\subsection{Model parameters}

The five parameters of the logarithmic PL potential \eqref{polyakov_potential} are listed in~{Table \ref{table1}.}
Originally, $T_0$ is the critical temperature of deconfinement for pure $SU(3)$ gauge theory, which is around $270$
MeV~\cite{Boyd,Kaczmarek}. Note that the chiral $T_c$ at zero density obtained in PNJL with $T_0=270$ MeV is quite
higher than the LQCD prediction~\cite{Laermann,Fodor_Katz_tem,Borsanyi:2010bp,Bazavov:2018mes}. Following~\cite{TS},
we adopt $T_0=195$ MeV here which can lead to a lower $T_c$.

The NJL part of PNJL has six parameters and a typical parameter set obtained in~\cite{Reinberg_Klevansky_Hufner,Klevansky}
is listed in {Table \ref{table2}}. These parameters are determined by the empirical values of $\eta'$ and
$\pi$ meson masses, the $\pi$ decay constant $f_\pi$, and the quark condensates at vacuum. To qualitatively
investigate the sensitivity of the RW transition on the $Z_3$ symmetry breaking patterns, we take the current quark
masses as free parameters in this study while keep $G_S$, $G_D$, and $\Lambda$ unchanged.

\begin{figure*}[!t]
 \centering
 \includegraphics[width=0.9\columnwidth]{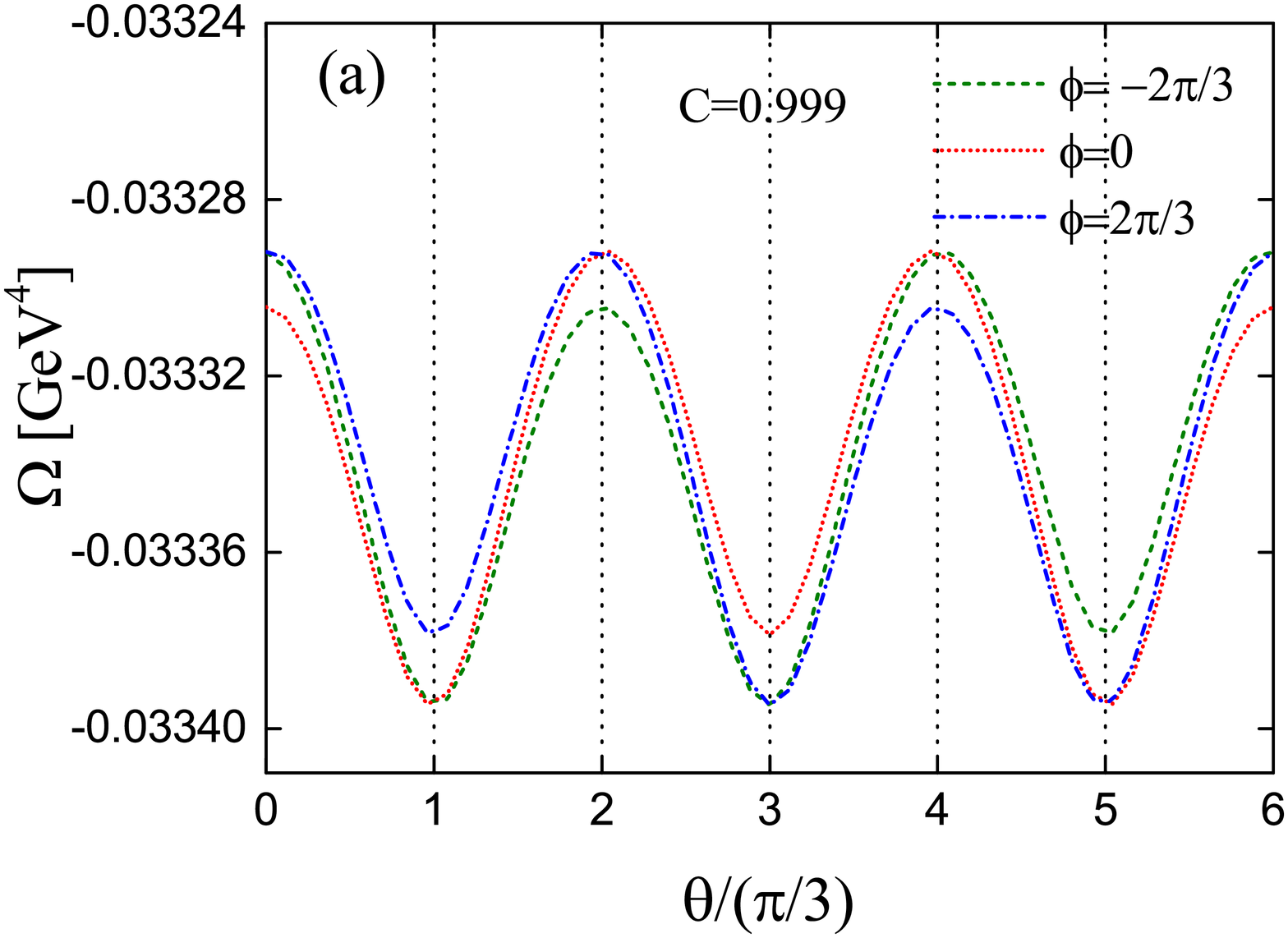}
 \includegraphics[width=0.9\columnwidth]{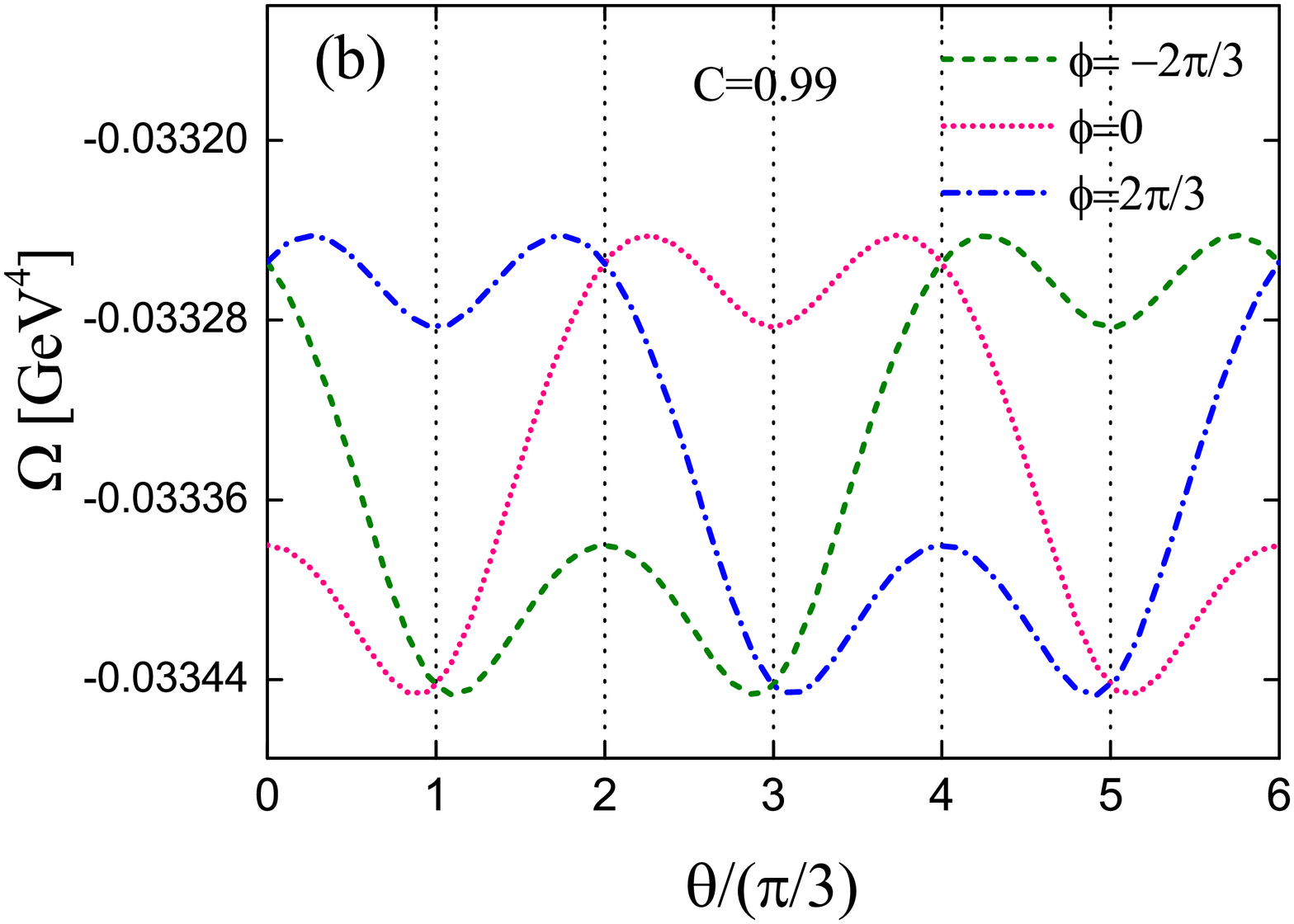}
 \includegraphics[width=0.9\columnwidth]{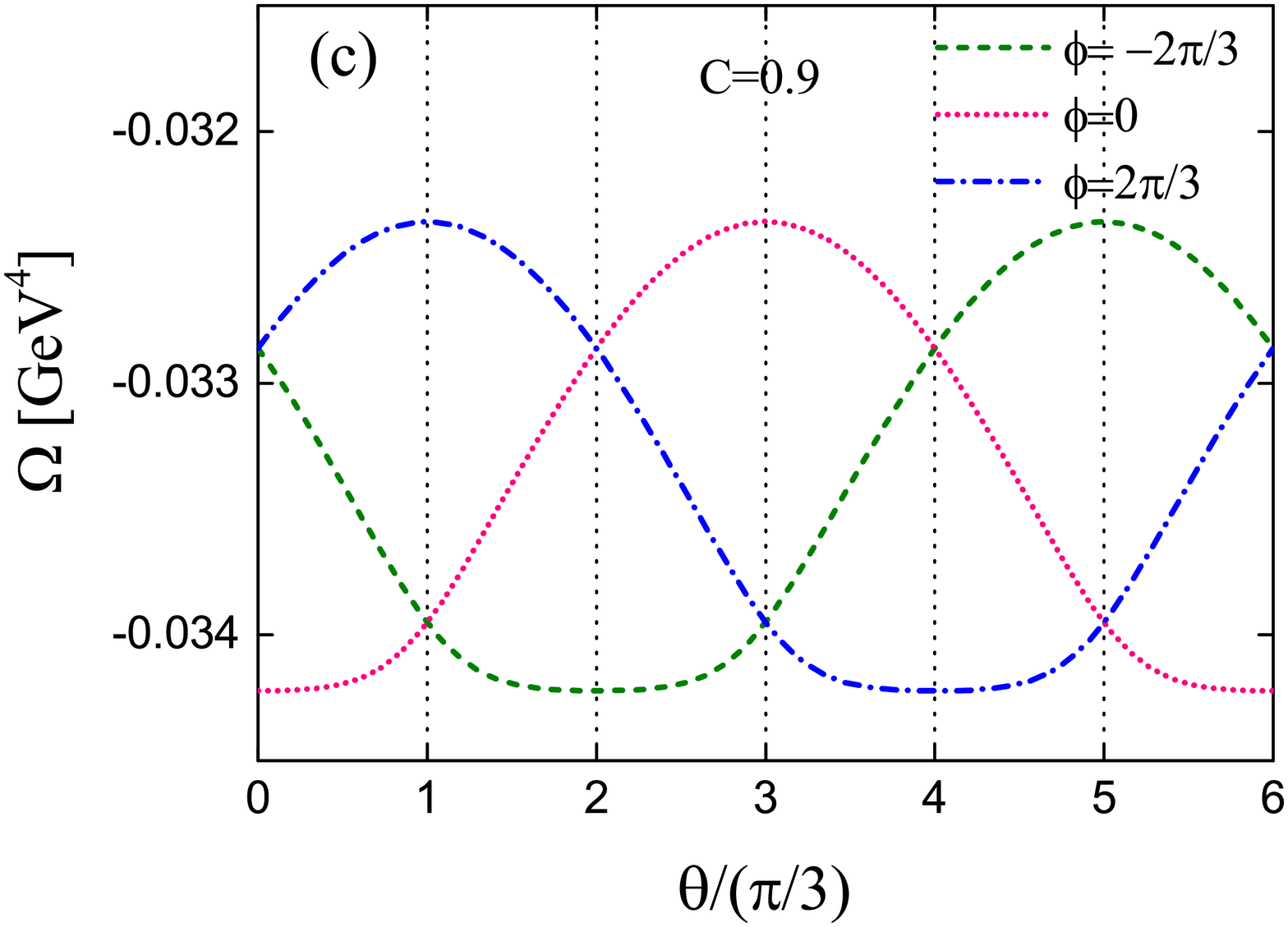}
 \includegraphics[width=0.9\columnwidth]{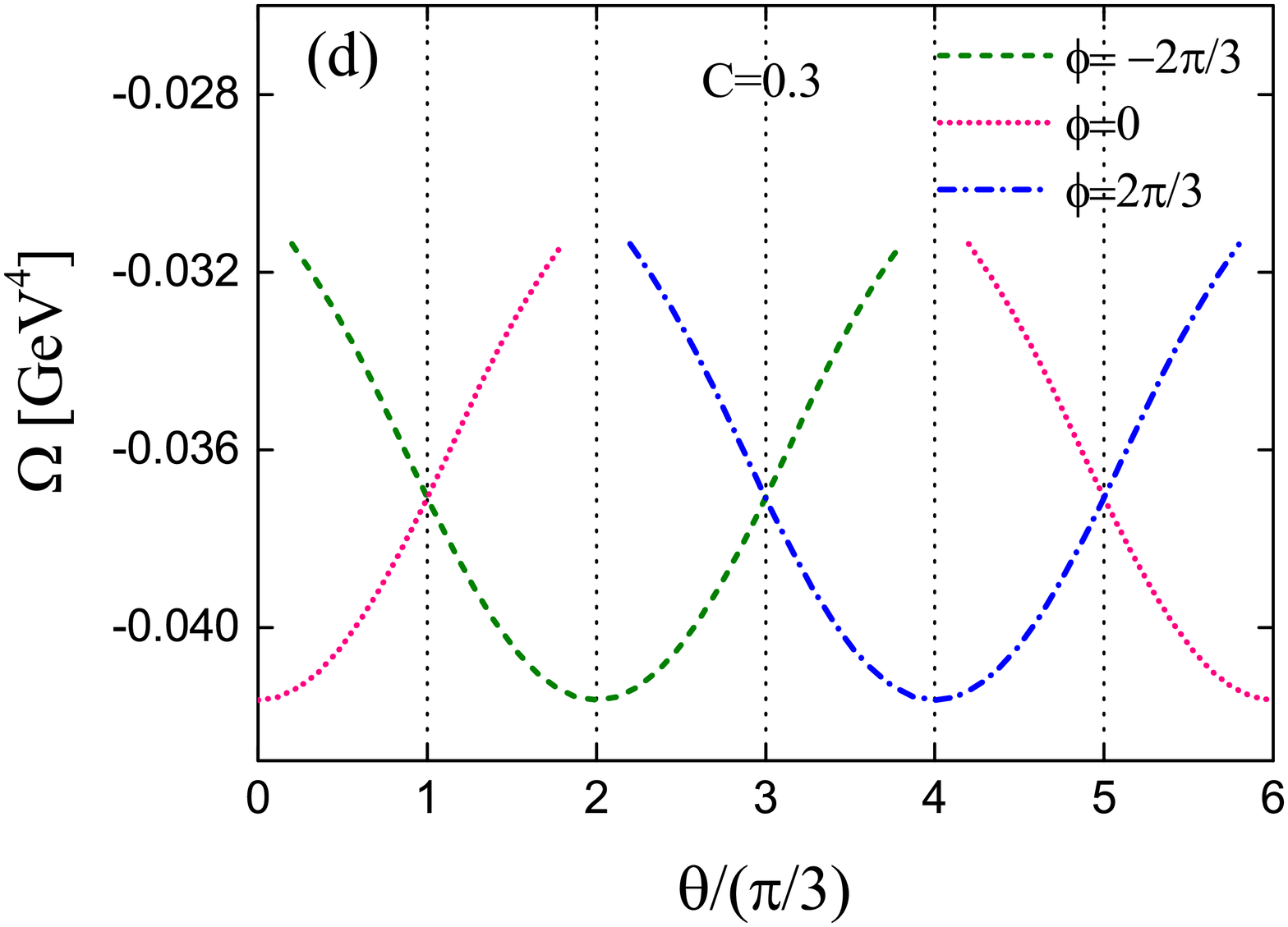}
 \caption{ Thermodynamic potentials of the $\mathbb{Z}_3$ sectors as the functions of $\theta$ at $T=250$
 MeV in PNJL for $N_f=3$ with fixed $m_u=m_d=m_s=5.5$ MeV and varied $C\neq{1}$.}
 \label{fig:2}
\end{figure*}

\begin{figure*}[htbp]
 \centering
 \includegraphics[width=0.9\columnwidth]{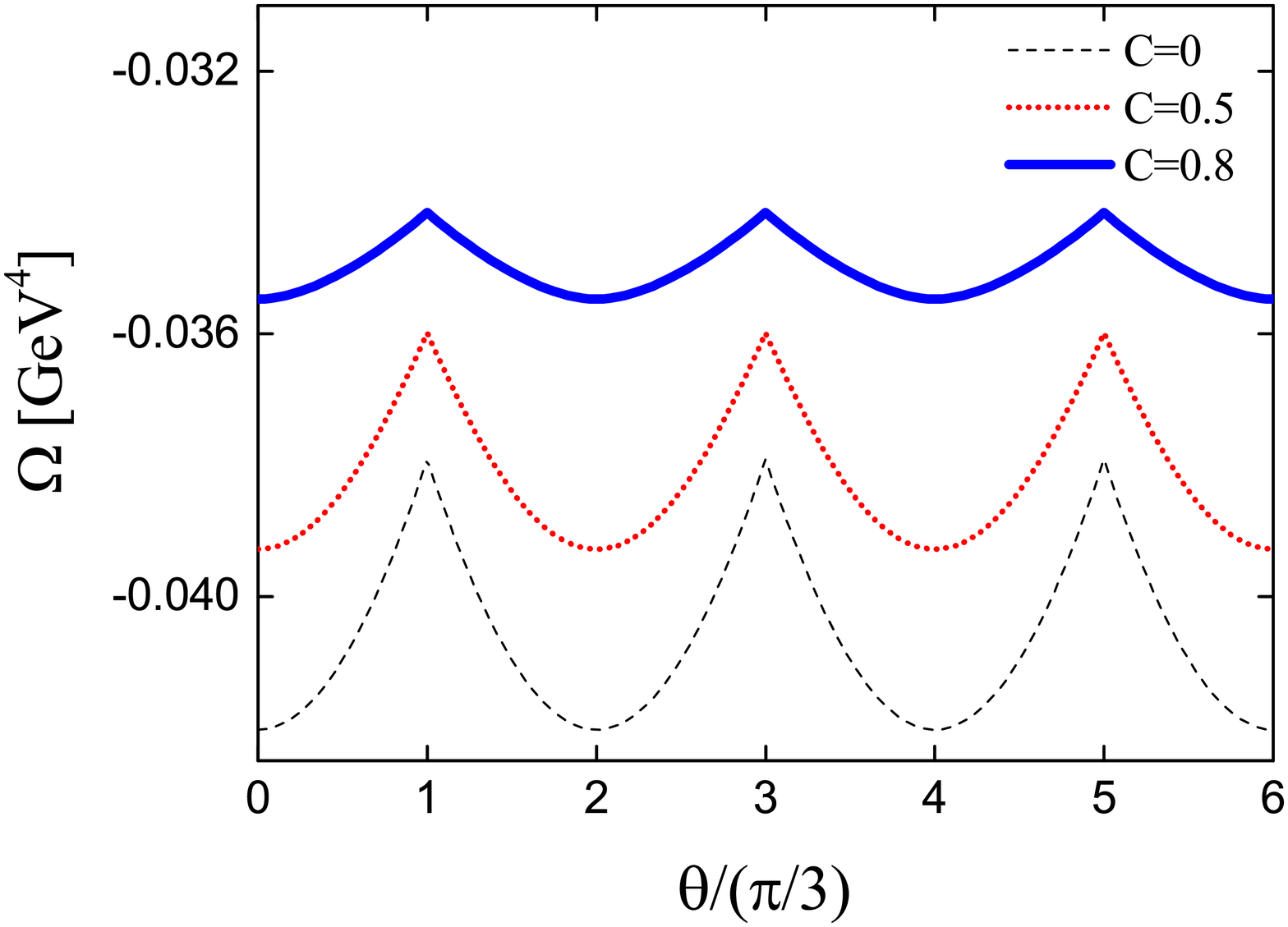}
 \includegraphics[width=0.9\columnwidth]{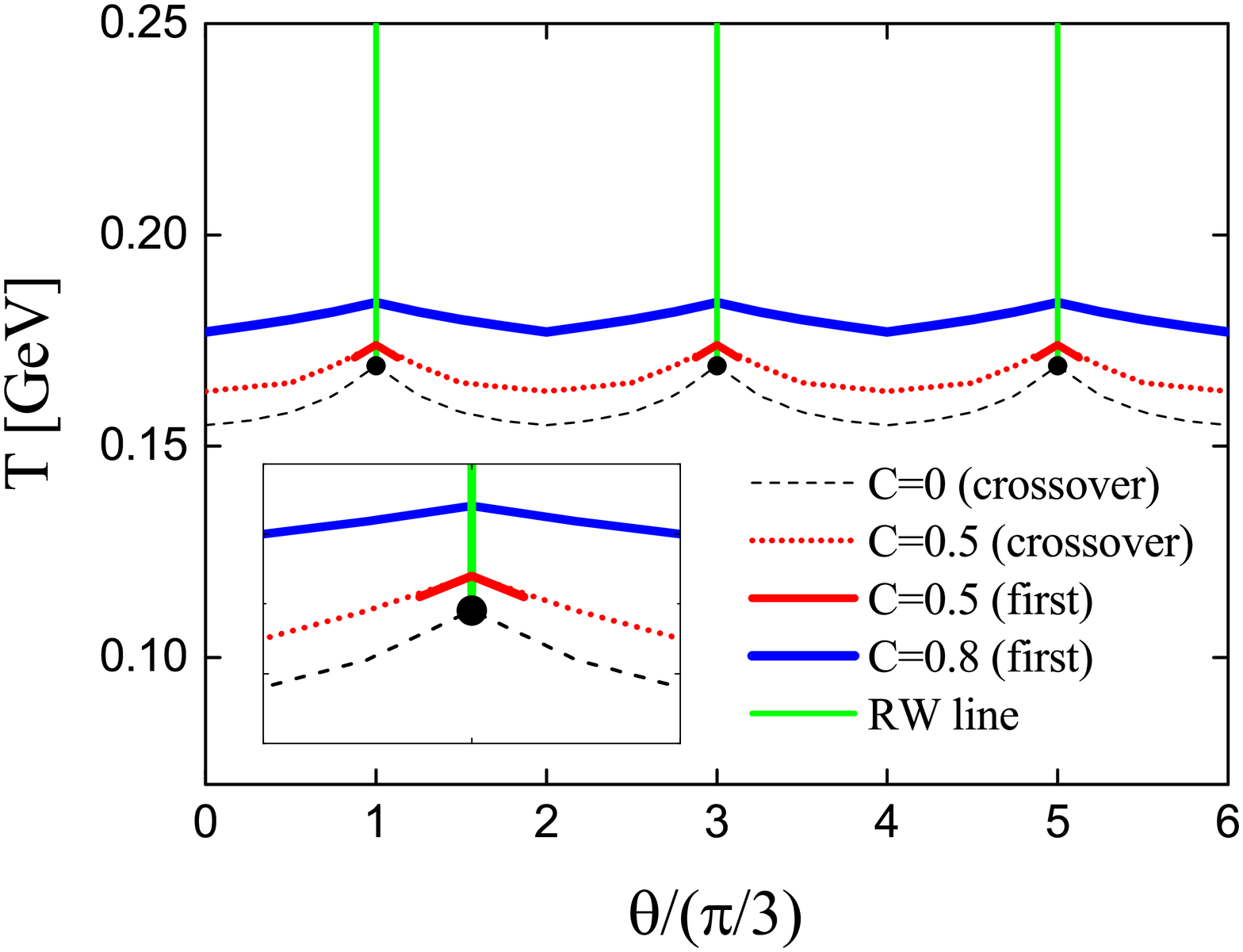}
 \caption{Thermodynamic potential $\Omega$ as the function of $\theta$ at $T=250$ MeV (left) and the $\theta$-T phase
 diagram (right) for $m_u=m_d=m_s=5.5$ MeV with varied $C\neq{1}$. The black spots in the right panel mean the RW endpoints
 at $C=0$ are still triple points.}
 \label{fig:3}
\end{figure*}

\section{Numerical results }
\label{Sec_3}
In this section, we show numerical results of PNJL with the imaginal chemical potentials
$(\mu_{u},\mu_{d},\mu_{s})/iT =(\theta-2C\pi/3, \theta, \theta+2C\pi/3)$, where $0\leq C\leq1$.
We study the RW and deconfinement transitions under the conditions (i)-(v) respectively. We concentrate
on how these transitions depend on the pattern of center symmetry breaking.

At high temperature, the thermodynamical potential of $\mathbb{Z}_3$-PNJL has three degenerate local minima
at $\phi=0$ and $\pm2\pi/3$, which are the three $\mathbb{Z}_3$ sectors. Correspondingly, the thermodynamic
potential of PNJL may have three non-degenerate solutions ( namely $\Omega_\phi$($\phi=0,\pm2\pi/3$)), and
the ground-state $\Omega_{gs}$ is determined by the absolute minimum of the three.

Without loss of generality, we take a fixed high temperature $T=250$ MeV to do the calculations at which
the RW transition always happens in this model.

\subsection{ Center symmetry breaking pattern (i): $N_f=3$ with varied $C\neq{1}$}

We first perform the calculation in $\mathbb{Z}_3$-PNJL with the small common quark mass $5.5~\text{MeV}$.
Fig.~\ref{fig:1} shows the thermodynamical potential $\Omega$ as the function of $\theta$ for two different
temperatures. We confirm that the ground state has a $3$-fold degeneracy at high temperature, and there is
no degeneracy at low temperature. The high-$T$ degeneracy indicates the spontaneous center symmetry breaking,
which rules out the RW transition. Numerical calculation indicates that the critical temperature 
$T_c\approx$ 195 MeV, which is almost independent on $\theta$~\cite{ZN1}. We see that the RW periodicity always 
holds even the $\theta$-dependence of $\Omega$ at low-$T$ ($<T_c$) is quite weaker than that at high-$T$ ($>T_c$). 
The upper panel displays that $\Omega(\theta)$ peaks at $\theta=(2k+1)\pi/3$ for $T=150$ MeV, but the lower 
shows it peaks at $\theta=2k\pi/3$ for $T=250$ MeV.  This implies that the $T$-driven first-order transition 
related to center symmetry corresponds to the shift of the shape of $\Omega(\theta)$. This is a nontrivial 
result in the center symmetric theory with fermions.

Figure.~\ref{fig:2} shows $\Omega_{\phi}(\theta)$ at $T=250$ MeV for $N_f=3$ with the same common quark mass
as in Fig.~\ref{fig:1} but $C\neq{1}$, which corresponds to center symmetry breaking pattern (i). We see the
shifts between three $\mathbb{Z}_3$ sectors appear in the $\theta-\Omega$ plane and the cusps of $\Omega$
emerge at $\theta=\theta_{rw}=(2k+1)\pi/3$. Note that the angel $\theta_{rw}$ is consistent with the traditional
one in QCD with $C=0$. Fig.~\ref{fig:2} displays that each $\Omega_\phi(\theta)$ has the period $2\pi$, which
is continuous (discontinuous) when center symmetry is weakly (strongly) broken for $C$ near one (zero).
Fig.~\ref{fig:2}(d) shows that the solution $\Omega_0(\theta)$ for $C=0.3$ vanishes in the region
$0.6\pi<\theta<1.4\pi$, which is similar to that of the standard RW transition obtained in the two-flavor
PNJL~\cite{Yahiro}. We notice that the PNJL correctly reproduces the relation $\Omega_{gs}(\theta)=\Omega_{gs}(-\theta)$
required by pattern (i). So the RW transitions shown in Fig.~\ref{fig:2} still reflect the spontaneous breaking
of $\mathbb{Z}_2$ symmetry and the density $\partial\Omega/\partial{(i\theta})$ is the order parameter.

\begin{figure}[!t]
 \centering
 \includegraphics[width=0.9\columnwidth]{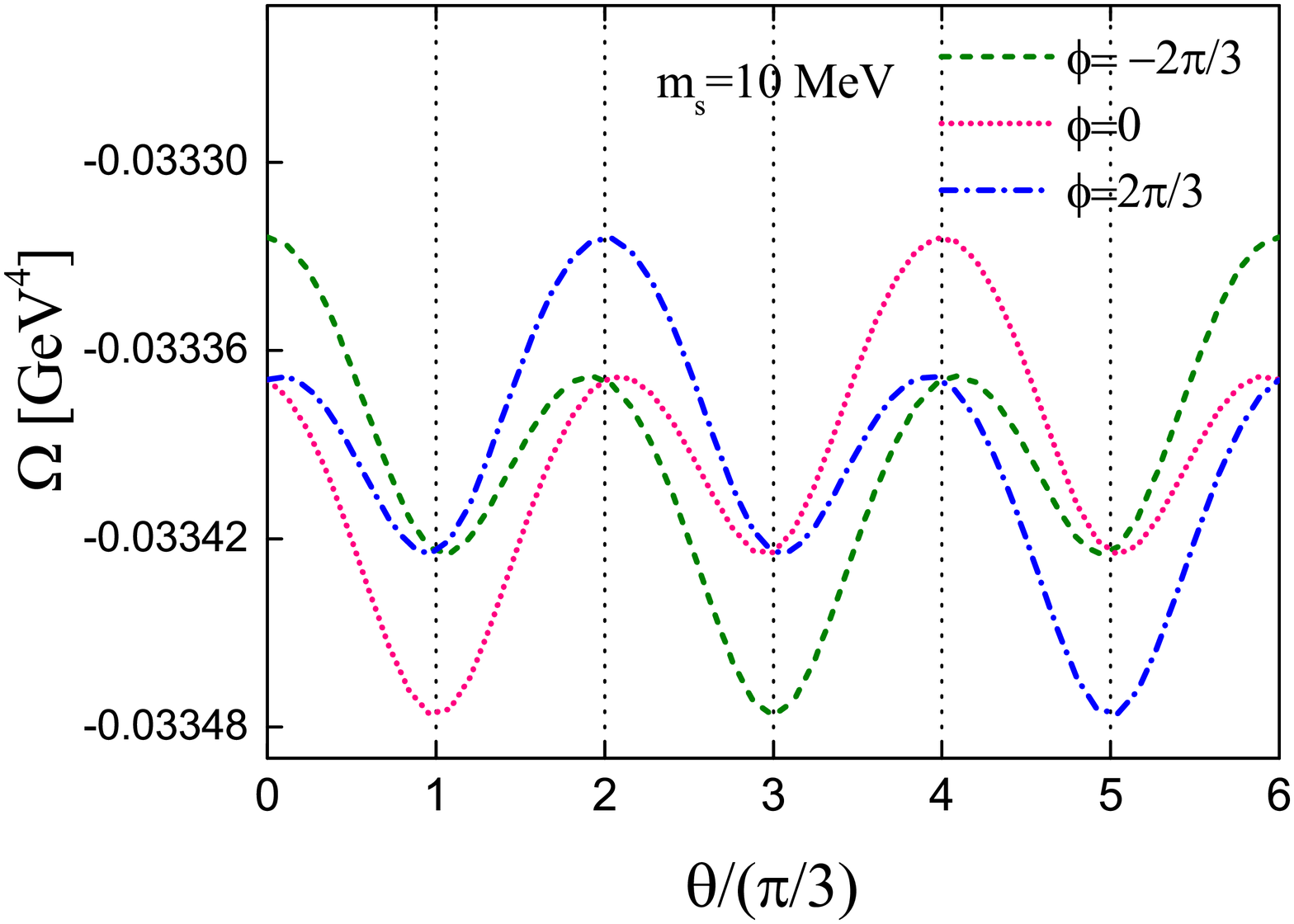}
 \includegraphics[width=0.9\columnwidth]{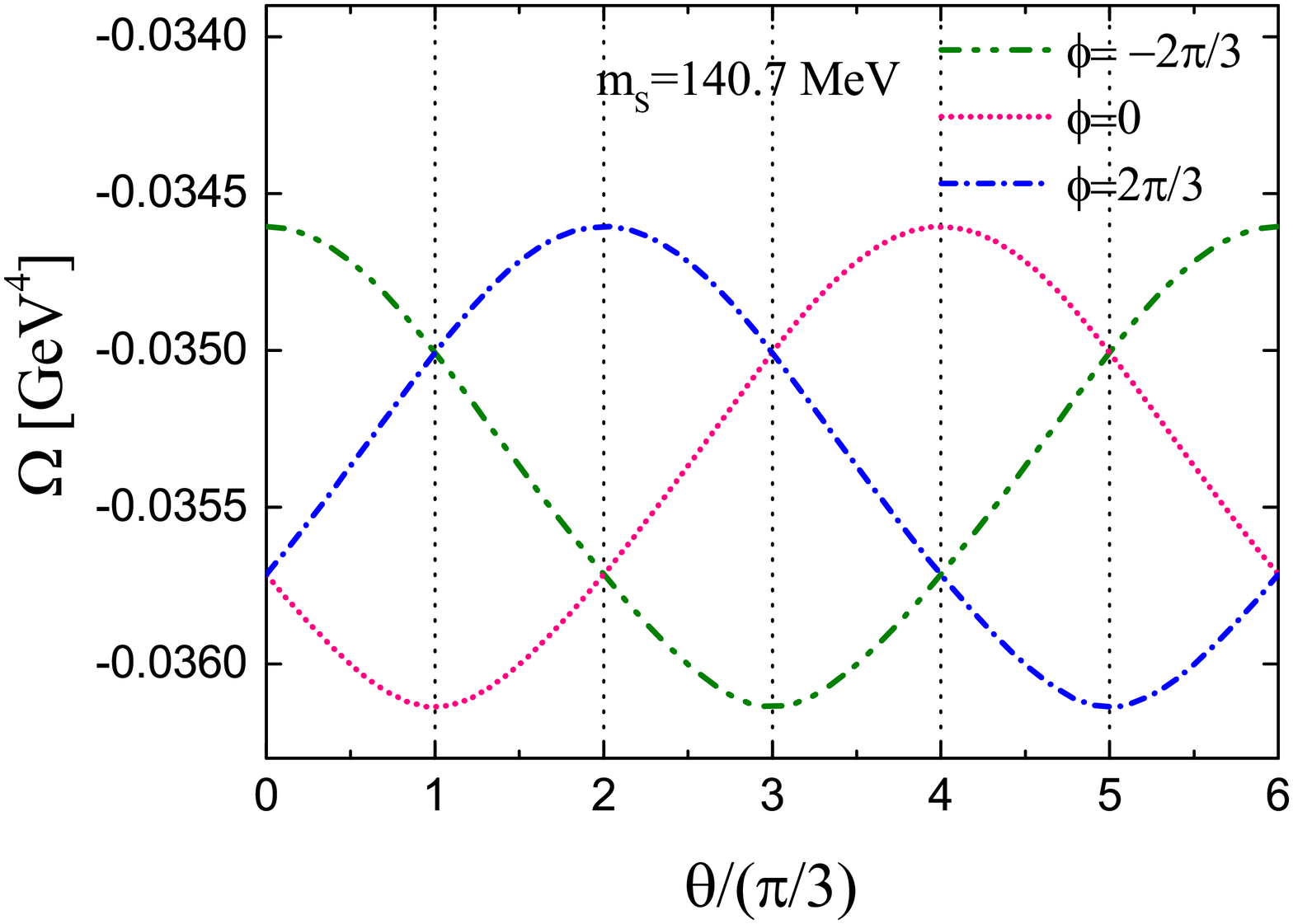}
 \caption{ Thermodynamic potentials of the $\mathbb{Z}_3$ sectors as the functions of $\theta$ at $T=250$
 MeV with $C=1$  for $m_s=10$ MeV (upper) and $m_s=140.7$ MeV (lower). The masses of two light flavors are
 fixed as $m_u=m_d=5.5$ MeV.  The RW transitions appear at $\theta=2k\pi/3$.}
 \label{fig:4}
\end{figure}

\begin{figure}[!t]
 \centering
 \includegraphics[width=0.9\columnwidth]{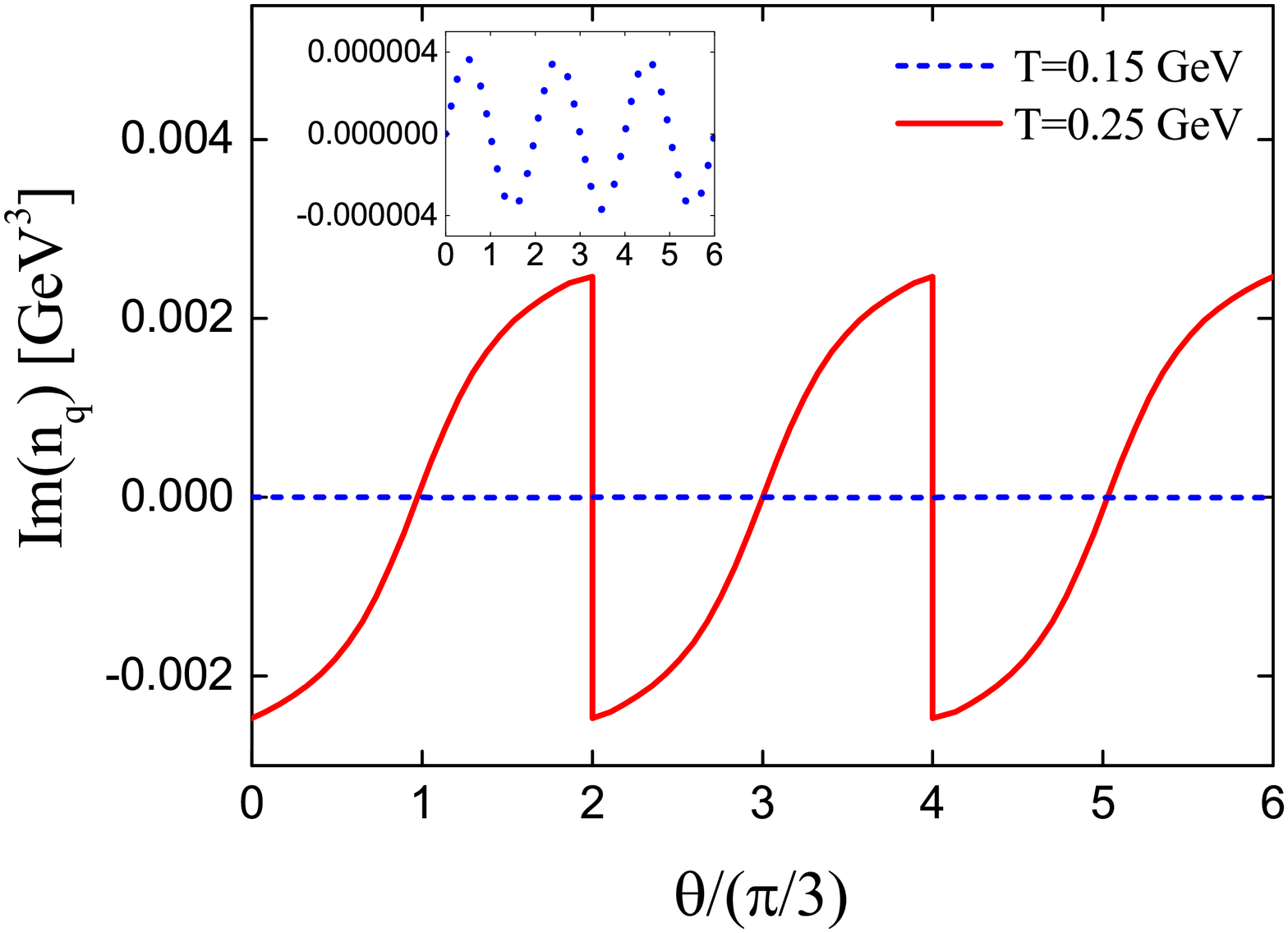}
 \caption{ The quark number density Im($n_q$) as a function of $\theta$ for $C=1$ at $T=150$
 MeV(dotted line) and $T=250$ MeV(solid line). The quark masses are same as that in the lower panel of Fig.~\ref{fig:4}.}
 \label{fig:5}
\end{figure}

\begin{figure}[!tb]
 \centering
 \includegraphics[width=0.9\columnwidth]{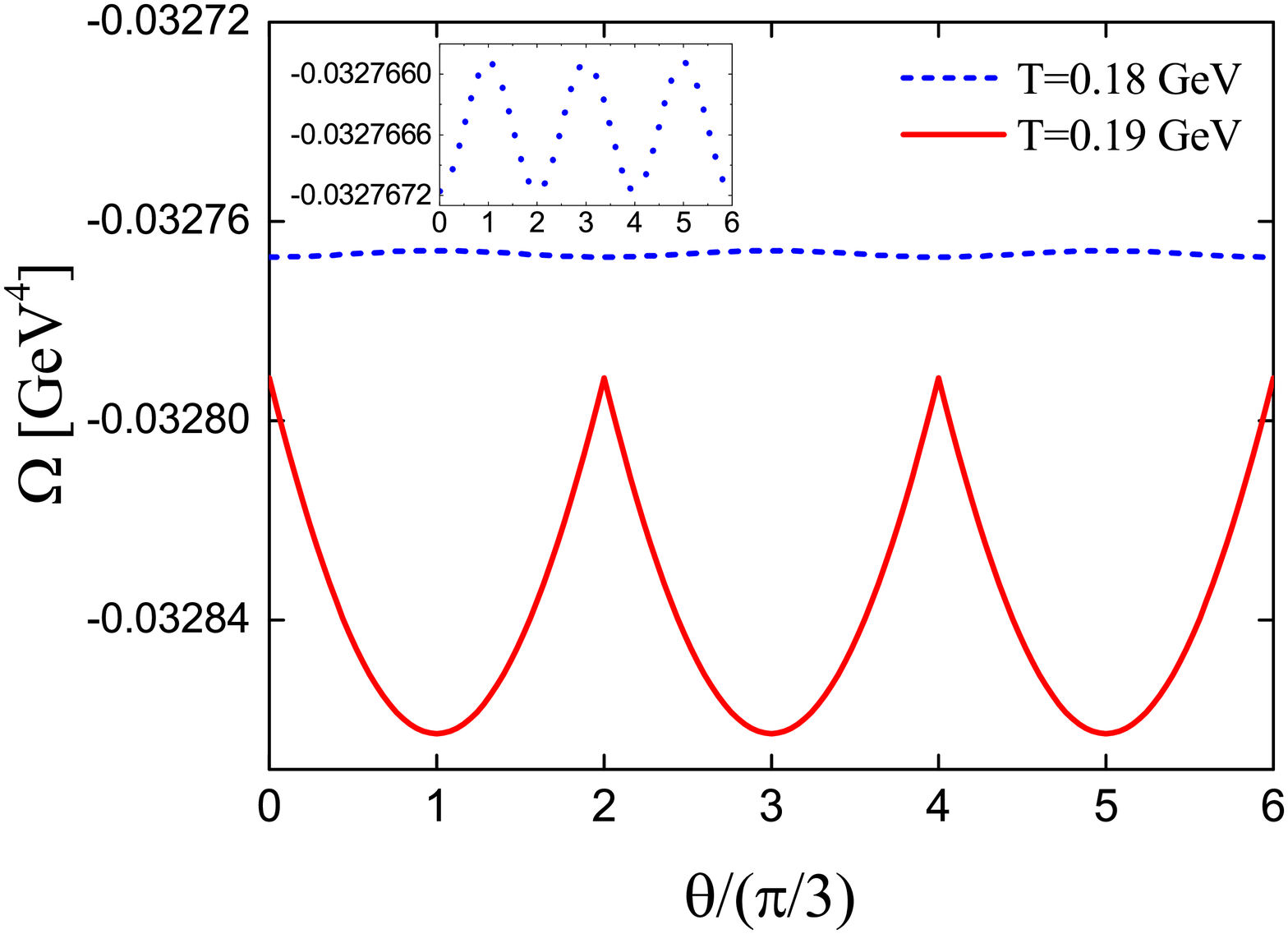}
 \caption{ Thermodynamic potential $\Omega$ as a function of $\theta$ for $C=1$ at $T=180$ MeV(dotted line)
 and $T=190$ MeV(solid line). The quark masses are same as that in the lower panel of Fig.~\ref{fig:4}.}
 \label{fig:6}
\end{figure}

As expected, Fig.~\ref{fig:2} shows that the RW cusps become sharper when $C$ declining from one to zero.
The RW transition getting stronger with center symmetry breaking is demonstrated more clearly in the left
panel of Fig.~\ref{fig:3}. In contrast, the deconfinement transition evaluated by the PL becomes weaker
with the decrease of $C$. This is shown in the right panel of Fig.~\ref{fig:3}, where the $\theta$-$T$
phase diagrams for three different $C's$ are plotted (the solid line denotes the first order transition).
In this panel, the vertical lines represent the RW transitions and the other lines the deconfinement transitions.
We see that for $C=0.8$, the whole line of deconfinement is first-order; but for $C=0.5$, only the short line near
the RW endpoint keeps first-order. When $C$ approaching zero, the first-order line of deconfinement further shrinks
towards the RW line but does not vanish at $C=0$. So all the RW endpoints for pattern (i) with a small quark mass
are triple points in this model
\footnote{ This is different from the current lattice predictions that the physical RW endpoint may be
second-order. Note that in PNJL with $C=0$, the nature of the RW endpoint depends on the PL potential.}.
Note that the triple point may change into a critical end point if the common quark mass is large enough.
In this case, there should exist a critical value $C_c$ below which the RW endpoint is second-order.

\subsection{Center symmetry breaking pattern (ii): $N_f=2+1$ with $C=1$}

This subsection gives the numerical results for $N_f=2+1$ and $C=1$, namely Pattern (ii), where the center
symmetry of $\mathbb{Z}_3$-QCD is broken by the mass difference between two degenerate light flavors (u and d)
and a heavy one (s).

\begin{figure}[!t]
 \centering
 \includegraphics[width=0.9\columnwidth]{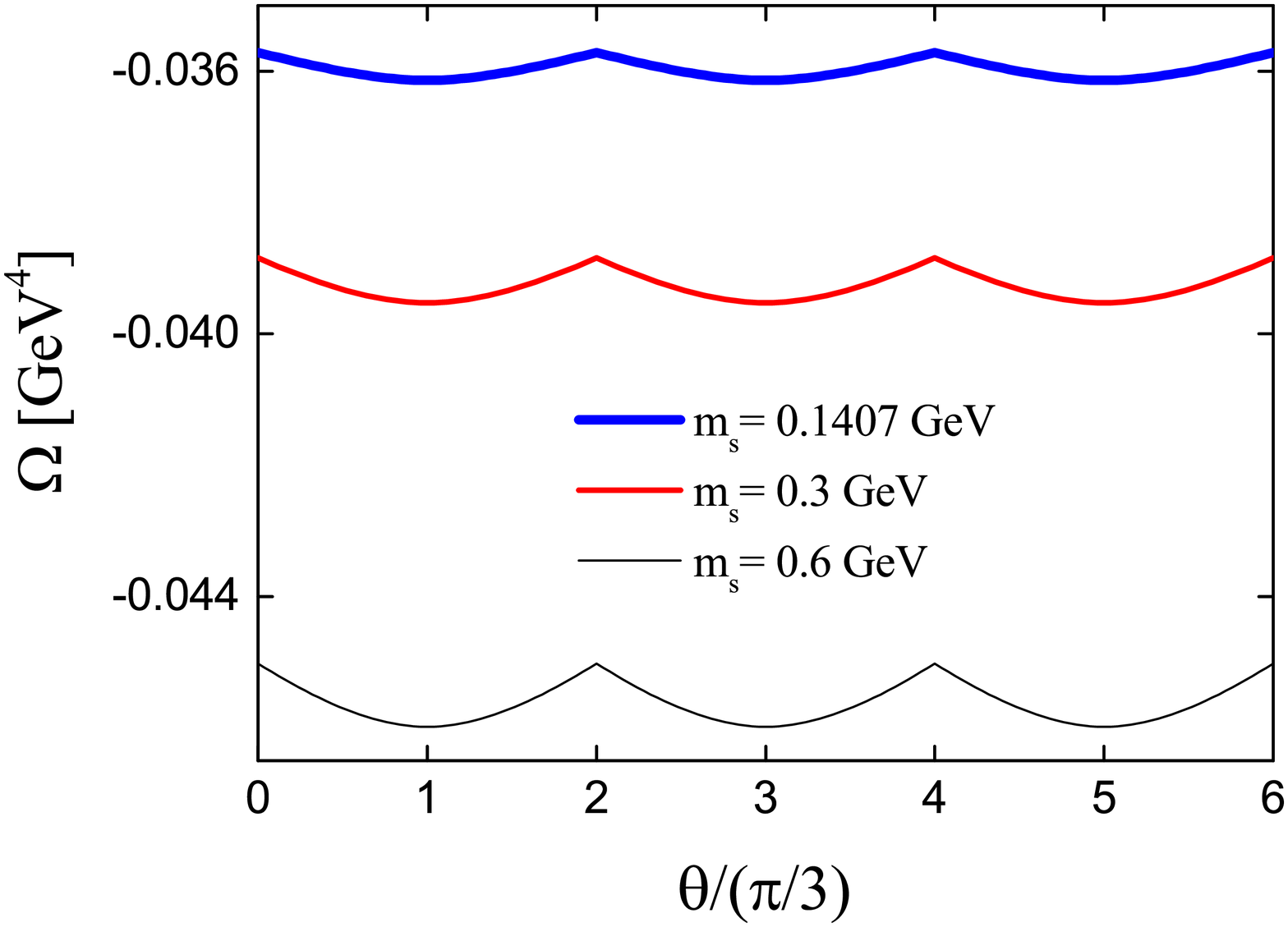}
 \includegraphics[width=0.9\columnwidth]{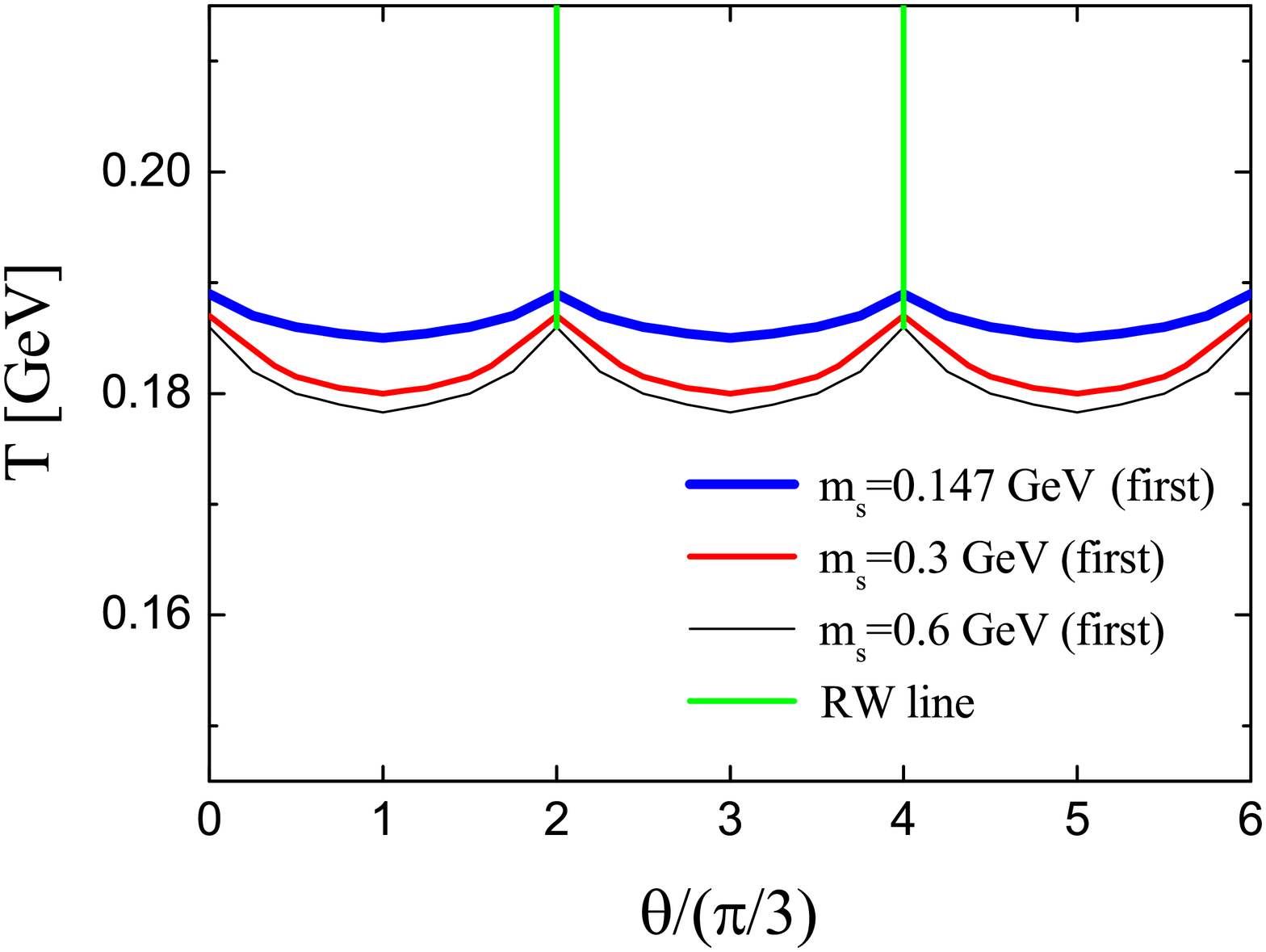}
 \caption{ The upper panel shows $\theta$-dependences of the thermodynamic potential $\Omega$ at $T=250$ MeV for
 $C=1$ and $m_u=m_d=5.5$ MeV with different strange quark masses. The lower panel shows the $\theta$-$T$ phase
 diagrams under the same conditions.}
\label{fig:7}
\end{figure}

Figure.~\ref{fig:4} presents the thermodynamic potential $\Omega_{\phi}$ as the function of $\theta$ at $T=250
$ MeV for two different $m_s$'s, where $m_{l(u,d)}=5.5$ MeV. In the range $0\leq\theta{<2\pi}$, each
$\Omega_{\phi}$ has three local minimums for $m_s=10$ MeV, but only one for $m_s=140.7$ MeV; the shifts between
three $\mathbb{Z}_3$ sectors appear in both cases. Similar to Pattern (i), the relation $\Omega(\theta)=\Omega(-\theta)$
is also reproduced corretly in PNJL. Different from Pattern (i), the RW cusps occur at $\theta=2k\pi/3$ rather
than $(2k+1)\pi/3$.

Note that $\theta_{RW}=2k\pi/3$ can be explained using the previous study \cite{Sakai:2009vb}, in which the
RW transitions at finite imaginal baryon and isospin chemical potentials, namely $\mu_{q(I)}=iT\theta_{q(I)}$,
are investigated in an $N_f=2$ PNJL. The prediction of \cite{Sakai:2009vb} is that:
(a) the RW transition emerges at $\theta_{q}=0$ mod $2\pi/3$ when $-\pi/2-\delta(T)<\theta_{I}<\pi/2+\delta(T)$
\footnote{Here $\delta(T)=0.00016\times{(T-250)}$.};
(b) it does at $\theta_{q}=\pi/3$ mod $2\pi/3$ when $\pi/2-\delta(T)<\theta_{I}<3\pi/2+\delta(T)$.
In our case with $N_f=2+1$ and $C=1$, $\theta_{q}$ and $\theta_{I}$ associated with two light flavors
are $((\theta-2\pi/3)+\theta)/2=\theta-\pi/3$ and $((\theta-2\pi/3)-\theta)/2=-\pi/3$, respectively.
According to (a) (ignoring the heavy flavor for the moment), $\theta_{I}=-\pi/3$ belongs to the range
$(-\pi/2-\delta(T),\pi/2+\delta(T))$, and thus the RW transition appears at $\theta_{q}=\theta_{RW}-\pi/3=\pi/3$
mod $2\pi/3$. Namely, $\theta_{RW}=2k\pi/3$. On the other hand, if we adopt
$(\mu_u, \mu_d, \mu_s)/iT=(\theta-2\pi/3, \theta+2\pi/3, \theta)$, the corresponding $\theta_{q}$ and $\theta_{I}$
are $\theta$ and $-2\pi/3$, respectively. In this case, $\theta_{I}+2\pi=4\pi/3$ belongs to the range
$(\pi/2-\delta(T),3\pi/2+\delta(T))$, and thus the RW transition occurs at $\theta_{q}=\theta_{RW}=0$ mod $2\pi/3$
according to (b). So we still obtain $\theta_{RW}=2k\pi/3$.

The consistency between our result and that in \cite{Sakai:2009vb} implies that the RW angle for $N_f=2+1$ with
$C=1$ is mainly determined by the two degenerate light flavors. Actually, we will show later that $\theta_{RW}$
is still $(2k+1)\pi/3$ for $N_f=1+2$ with $C=1$, in which there is only one light flavor.

\begin{figure}[!t]
 \centering
 \includegraphics[width=0.9\columnwidth]{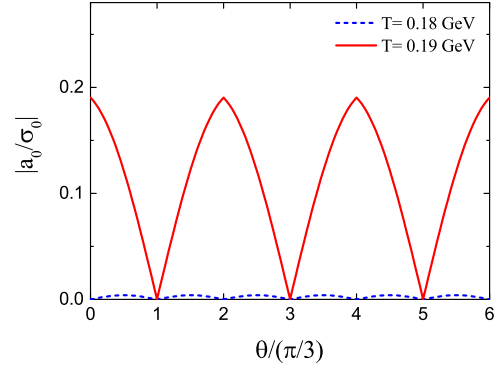}
 \caption{The isovector condensate $a_0$ as a function of $\theta$ for $C=1$ at $T=180$ MeV(dotted line) and
 $190$ MeV(solid line). The quark masses are same as that in the lower panel of Fig.~\ref{fig:4}. }
\label{fig:8}
\end{figure}

Figure.~\ref{fig:5} displays the $\theta$-dependence of the quark number density $n_I$=Im($n_q$) for
$m_{l(u,d)}=5.5$ MeV and $m_s=140.7$ MeV. In Pattern (ii), $n_I(\theta)$ is $\theta$-odd and thus the order
parameter for $\mathbb{Z}_2$ symmetry. We see that it is continuous for $T=150$ MeV but discontinuous
at $\theta=2k\pi/3$ for $T=250$ MeV.

In Fig.~\ref{fig:6}, we compare the thermodynamic potentials for temperatures below and above $T_{RW}$.
For $T=180$ MeV ($<T_{RW}$), $\Omega$ is weakly dependent on $\theta$, of which peaks and troughs are located
at $\theta=(2k+1)\pi/3$ and $2k\pi/3$, respectively; but for $T=190$ MeV ($>T_{RW}$), it depends on $\theta$
obviously and the positions of peak and trough are exchanged. Note that the peak and trough locations of
$\Omega$ at low and high temperature in Fig.~\ref{fig:6} are all the same as that of $\mathbb{Z}_3$-PNJL
shown in Fig.~\ref{fig:1}.
This indicates that the center symmetry is broken weakly in Pattern (ii) with the physical quark masses.

\begin{figure}[!t]
 \centering
 \includegraphics[width=0.9\columnwidth]{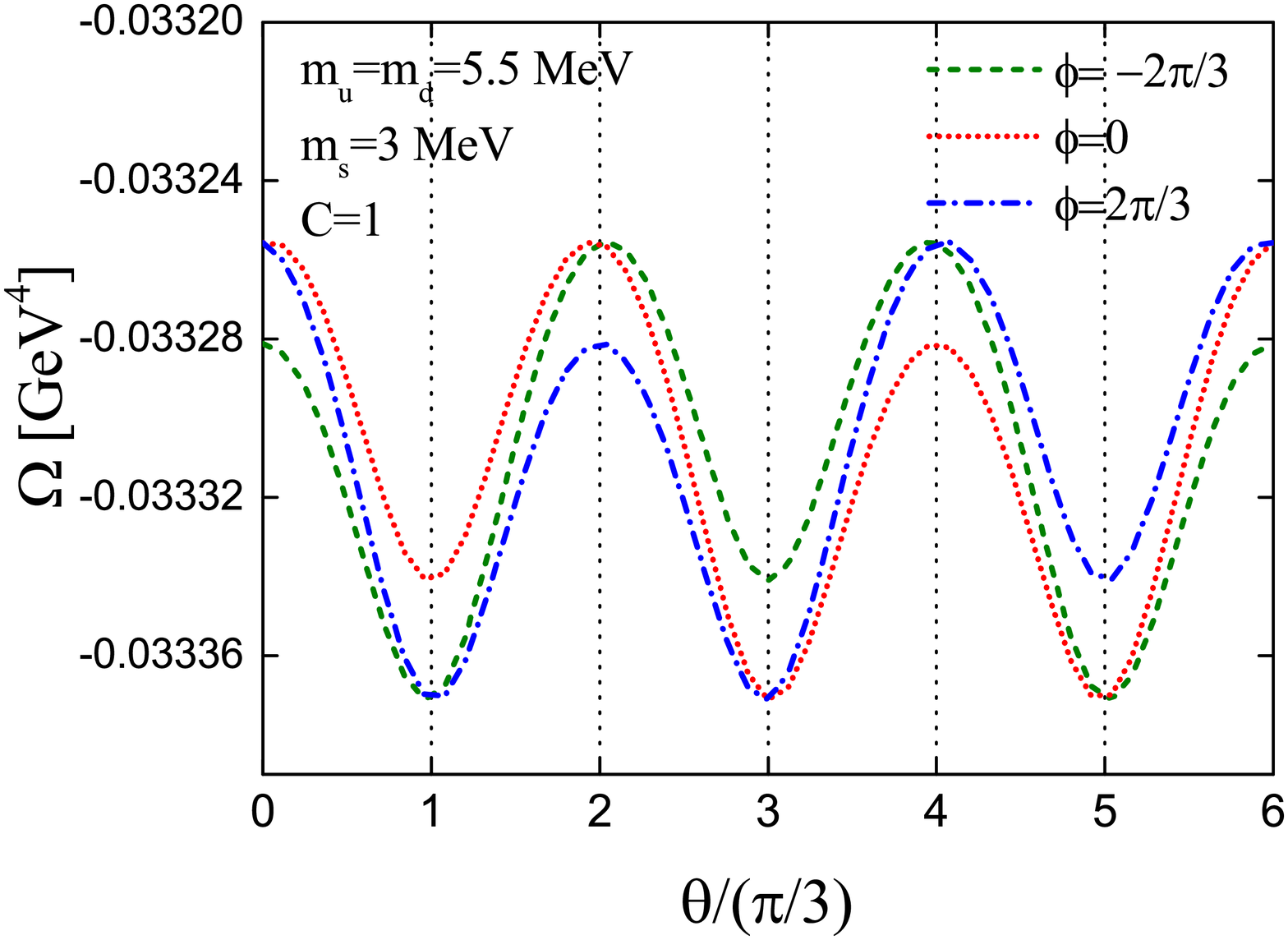}
 \includegraphics[width=0.9\columnwidth]{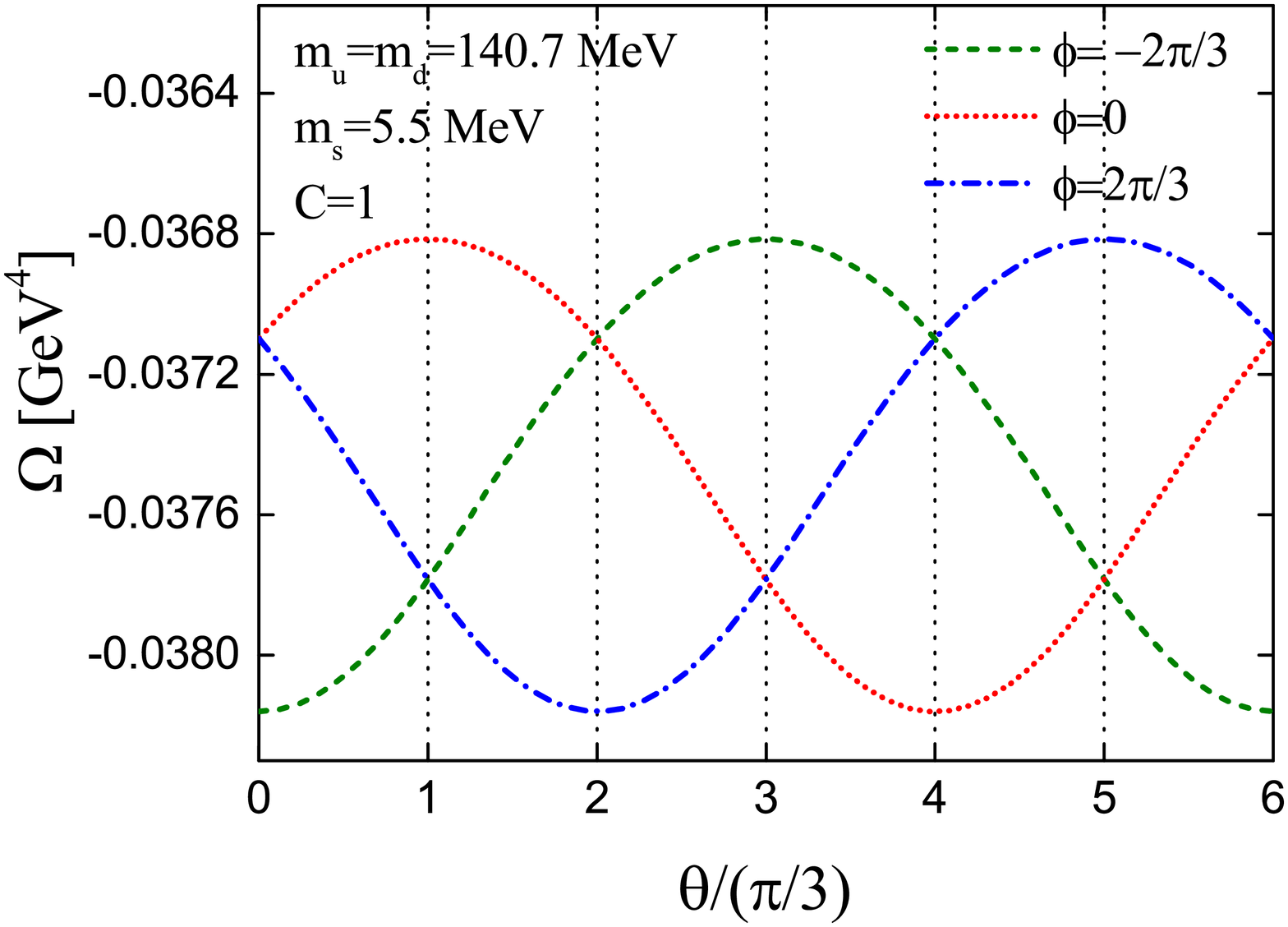}
 \caption{ Thermodynamic potentials of the $\mathbb{Z}_3$ sectors as the functions of $\theta$ at $T=250$ MeV
 for $N_f=1+2$ with $C=1$. The upper (lower) panel corresponds to the case with $m_u=m_d=5.5$ MeV and $m_s=3$
 MeV ($m_u=m_d=140.7$ MeV and $m_s=5.5$ MeV). The RW transitions appear at $\theta=(2k+1)\pi/3$.}
 \label{fig:9}
\end{figure}

The upper panel of Fig.~\ref{fig:7} presents thermodynamic potentials as functions of $\theta$ at $T=250$ MeV
for different larger $m_s$'s ($m_s=140.7,300,600$ MeV) with fixed $m_{l}=5.5$ MeV. As expected, the RW cusps
become sharper with $m_s$, but the change is milder (We only consider $m_s < \Lambda$ due to the limitation of PNJL).
The lower panel of Fig.~\ref{fig:7} displays the $\theta$-$T$ phase diagrams under the same conditions.
The deconfinement transitions for those $m_s$'s are all first-order, and thus the RW endpoints are triple points.
This suggests that the center symmetry is broken less severely by the mass differences considered here.
Similar to Fig.~\ref{fig:3}, the lower panel shows the higher the degree of center symmetry breaking, the lower
the $T_{RW}$. Another common feature of Fig.~\ref{fig:3} and Fig.~\ref{fig:7} is that $T_{RW}$ is the highest critical
temperature of deconfinement for a fixed $C\neq{1}$ (former) or $m_s$ (later). This implies that the RW transition
has a significant impact on the deconfinement transition in both symmetry breaking patterns.

In Fig.~\ref{fig:8}, we show the isovector condensate $a_0=\left\langle\bar{u}u-\bar{d}d\right\rangle$ as a function
of $\theta$ for $C=1$ with the physical quark masses. Here $a_0$ is normalized by $\sigma_0\equiv \sigma(T=0,\mu_f=0)$,
where $\sigma\equiv \left(\sigma_u+\sigma_d+\sigma_s\right)/3$. For $T=180$ MeV, the $a_0\thicksim0$ apart from
$\theta=k\pi/3$ where $a_0=0$. This can be considered as a remanent of the $N_f=3$ case where $a_0=0$ at low-$T$ \cite{ZN1}.
The approximate flavor symmetry at low-$T$ is attributed to the color confinement where $\Phi\sim0$. When $\theta=k\pi/3$,
the charge conjugate symmetry preserves the flavor symmetry between u and d. Actually, under the $\mathcal{C}$ transformation
$\Phi\leftrightarrow\Phi^*$, the thermodynamic potential with $\theta=0$
\begin{align}
 &\Omega(-2\pi/3,0,2\pi/3)\xrightarrow{\mathcal{C}}\Omega(2\pi/3,0,-2\pi/3)\xlongequal{-2\pi/3}\notag \\
 &\Omega(0,-2\pi/3,-4\pi/3)\xlongequal[u\leftrightarrow d]{-4\pi/3\rightarrow 2\pi/3}\Omega(-2\pi/3,0,2\pi/3),
\end{align}
and that with $\theta=\pi/3$
\begin{eqnarray}
&\Omega(-\pi/3,\pi/3,\pi)\xrightarrow{\mathcal{C}}\Omega(\pi/3,-\pi/3,-\pi)\xlongequal[u\leftrightarrow d]{-\pi\rightarrow \pi}\notag \\
 &\Omega(-\pi/3,\pi/3,\pi),
\end{eqnarray}
where $u\leftrightarrow d$ stands for the relabeling of u and d. For $T=190$ MeV, the two flavor symmetry
is broken at $\theta=2k\pi/3$ due to the RW transition\cite{ZN1,Yahiro}.

\subsection{Center symmetry breaking pattern (iii): $N_f=1+2$ with $C=1$}

The numerical results for $N_f=1+2$ and $C=1$ or Pattern (iii) are given in this subsection. The center
symmetry is broken by the mass difference between the light flavor and two degenerate heavy ones.
Note that s refers to the only light flavor here.

\begin{figure}[!t]
 \centering
 \includegraphics[width=0.9\columnwidth]{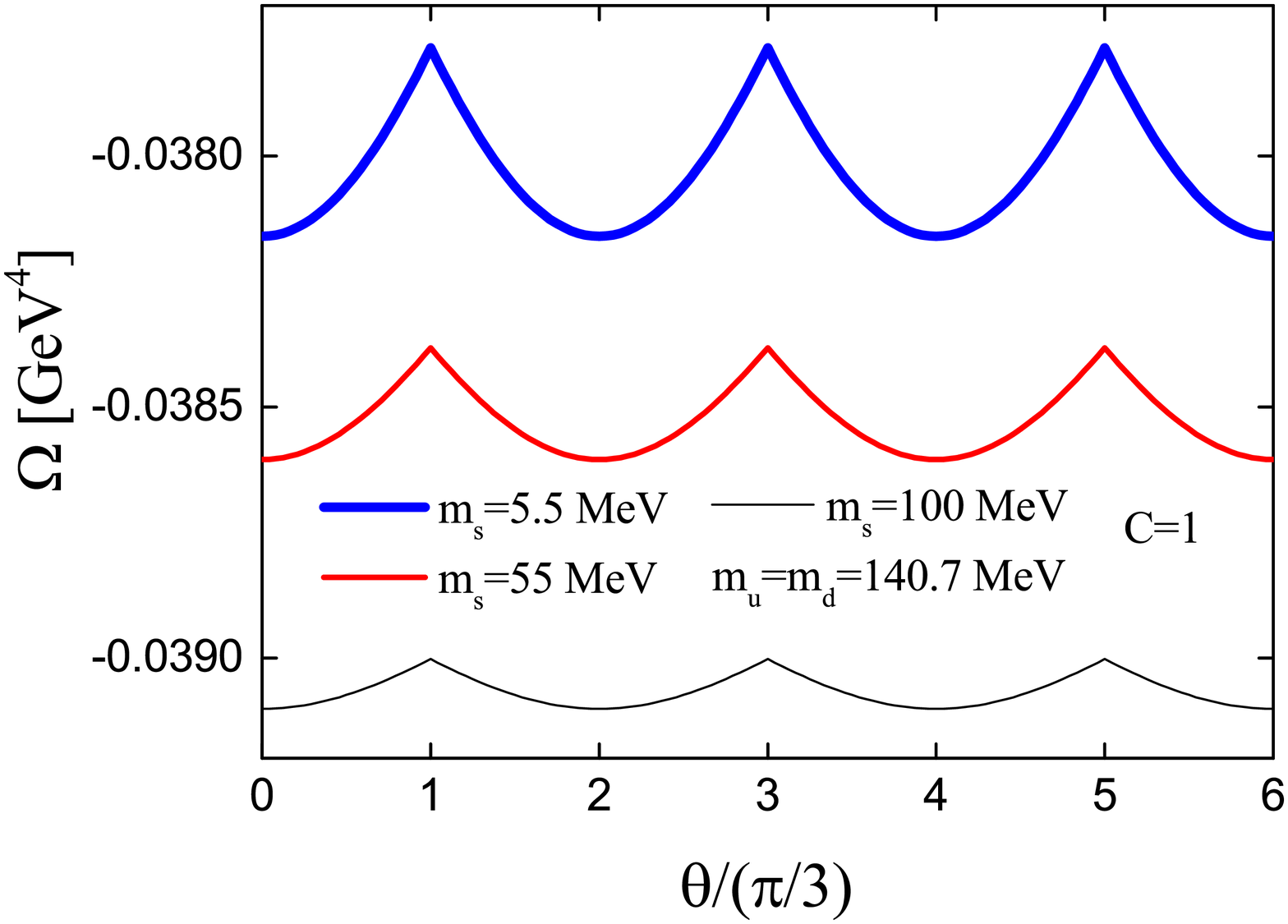}
 \includegraphics[width=0.9\columnwidth]{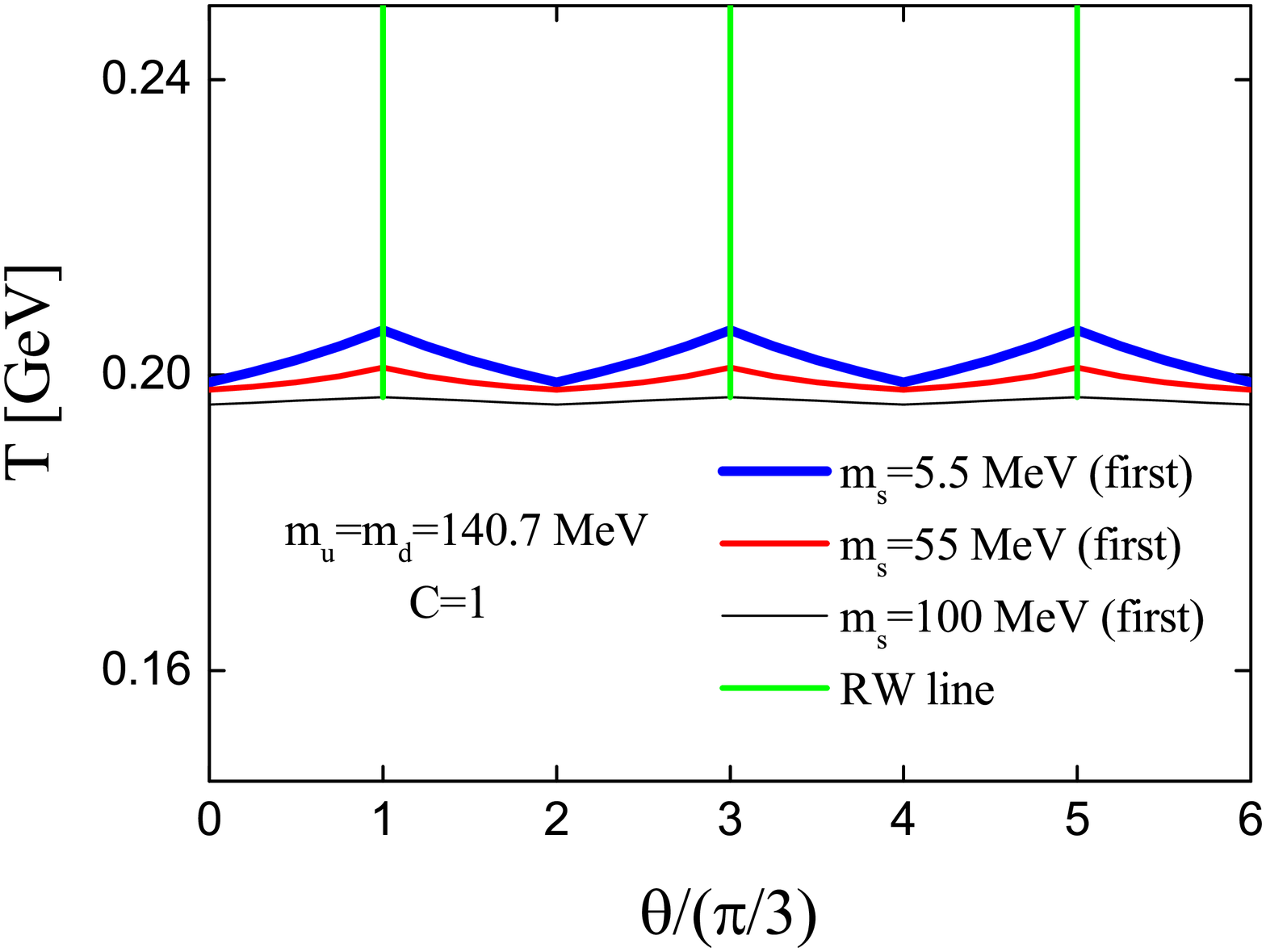}
 \caption{ The upper panel shows $\theta$-dependences of the thermodynamic potential $\Omega$ at $T=250$ MeV for different
 $m_s$'s and fixed $C=1$, where $m_u=m_d=140.7$ MeV. The lower one shows the $\theta$-$T$ phase diagrams under the same conditions.}
\label{fig:10}
\end{figure}

Figure.~\ref{fig:9} shows $\mathbb{Z}_3$ sectors of $\Omega$ as functions of $\theta$ at
$T=250~\text{MeV}$ for two cases with $m_{u(d)}=5.5$ MeV and $m_s=3$ MeV (upper panel) and
$m_{u(d)}=140.7$ MeV and $m_s=5.5$ MeV (lower panel). Different from Pattern (ii), both
panels display that RW transitions occur at $\theta=(2k+1)\pi/3$. This difference can be
understood in the following way. In Fig.~\ref{fig:9}, the physical thermodynamic potentials in
intervals $-\frac{\pi}{3}<\theta<\frac{\pi}{3}$, $\frac{\pi}{3}<\theta<\pi$, and
$\pi<\theta<\frac{5\pi}{3}$ are $\Omega_{\phi=-\frac{2\pi}{3}}$, $\Omega_{\phi=\frac{2\pi}{3}}$,
and $\Omega_{\phi=0}$, respectively. Note that such an $\Omega_{\phi}$ order of the physical
thermodynamic potential along the $\theta$ direction is same as that of one flavor system with
$\mu=\mu_{s}=i(\theta+\frac{2\pi}{3})T$
\footnote{The $\Omega_{\phi}$ order of the thermodynamic potential for one flavor system with
$\mu/iT=\theta$ is $\Omega_{\phi=0}$, $\Omega_{\phi=-\frac{2\pi}{3}}$, and $\Omega_{\phi=\frac{2\pi}{3}}$
for the $\theta$ intervals mentioned above \cite{Weiss}. When $\mu/iT=\theta+\frac{2\pi}{3}$, the
thermodynamic potential is shifted by $-\frac{2\pi}{3}$ along the $\theta$ axis and thus the order
becomes $\Omega_{\phi=-\frac{2\pi}{3}}$, $\Omega_{\phi=\frac{2\pi}{3}}$, and $\Omega_{\phi=0}$.}.
This suggests that $\theta_{rw}$ for $N_f=1+2$ and $C=1$ is mainly determined by the only light flavor.
Such a conclusion also supports our argument that $\theta_{rw}$ for $N_f=2+1$ and $C=1$ is mainly
determined by the two degenerate light flavors.

\begin{figure*}[!t]
\centering
 \includegraphics[width=0.9\columnwidth]{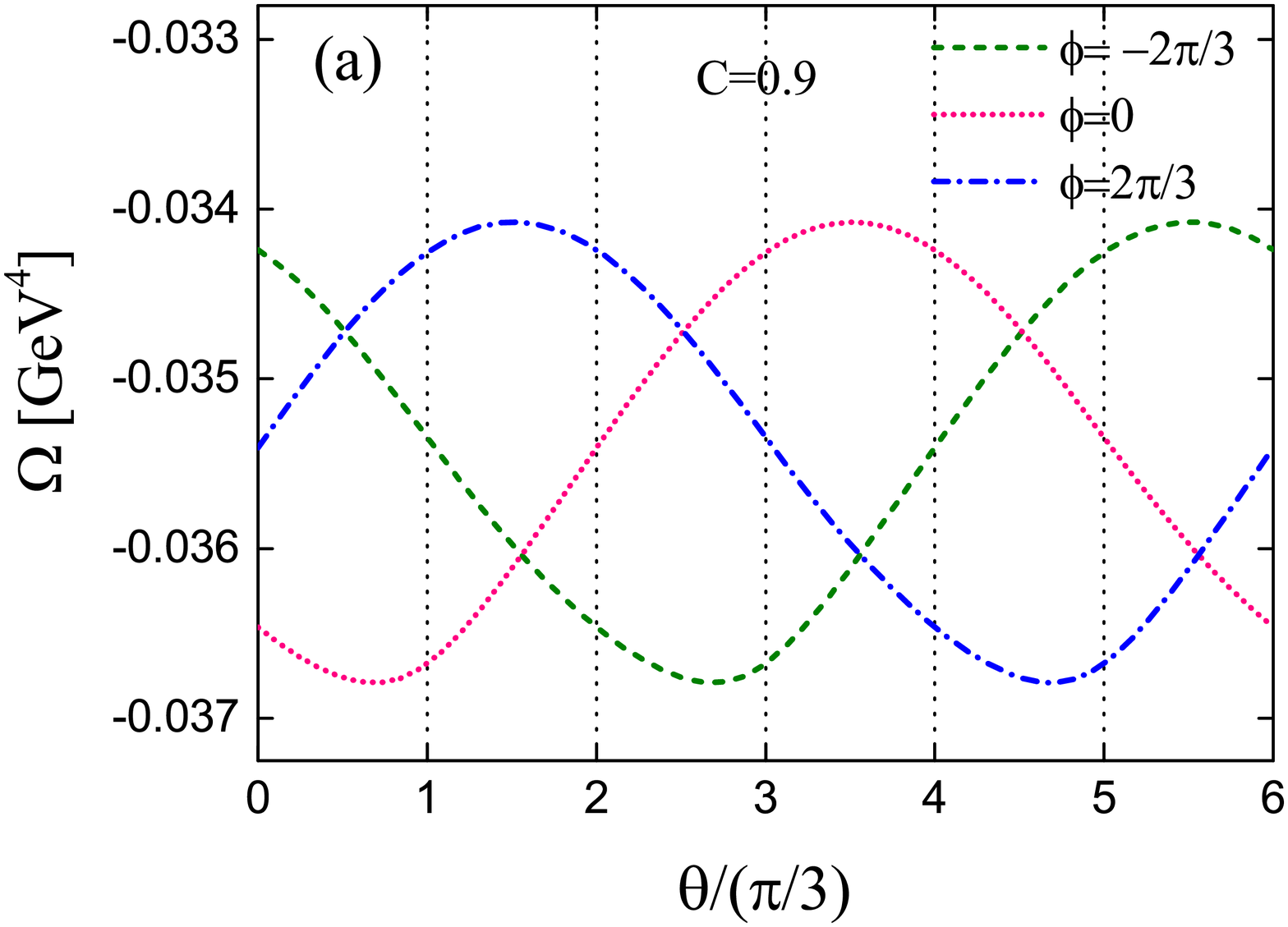}
 \includegraphics[width=0.9\columnwidth]{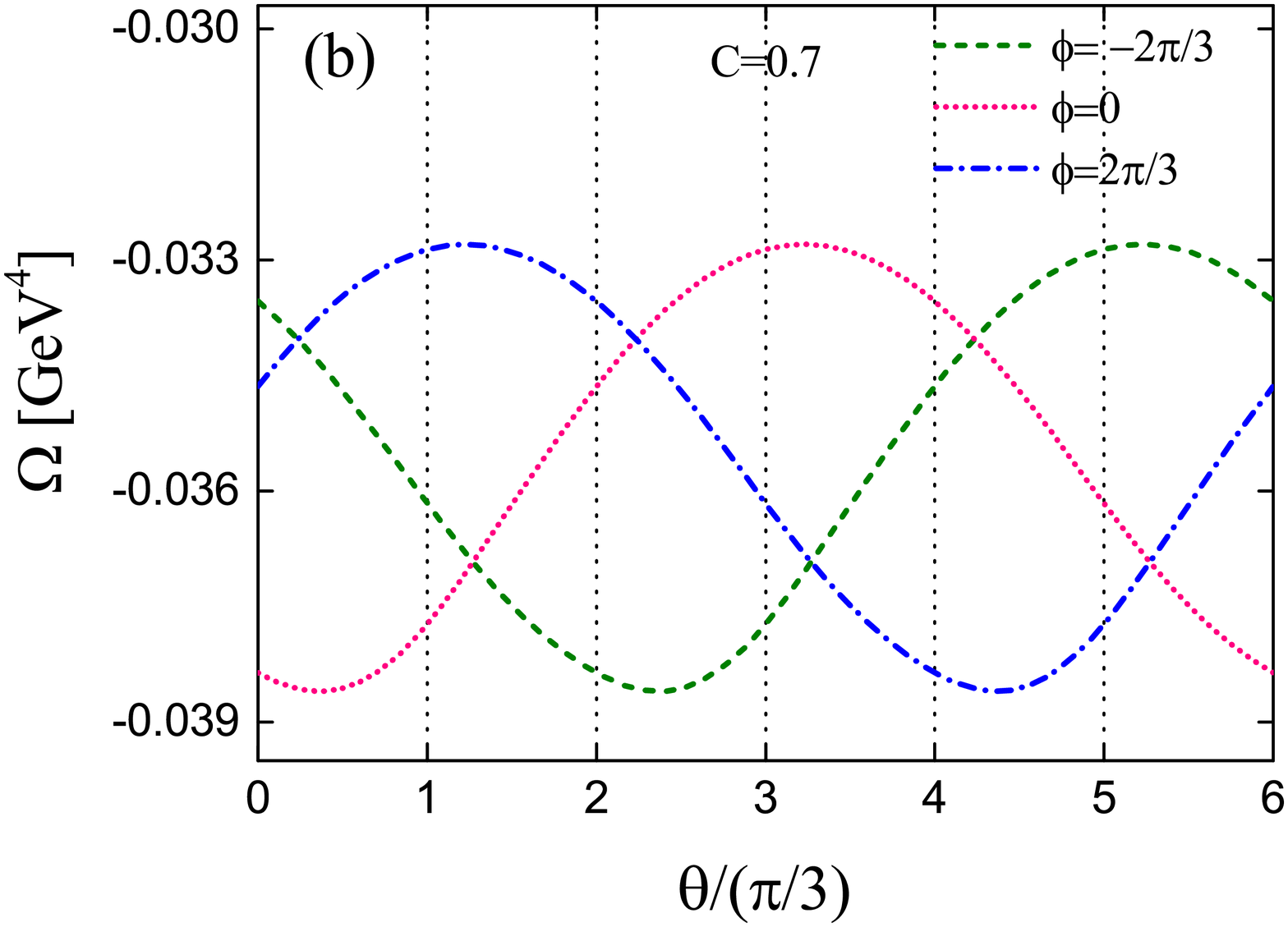}
 \includegraphics[width=0.9\columnwidth]{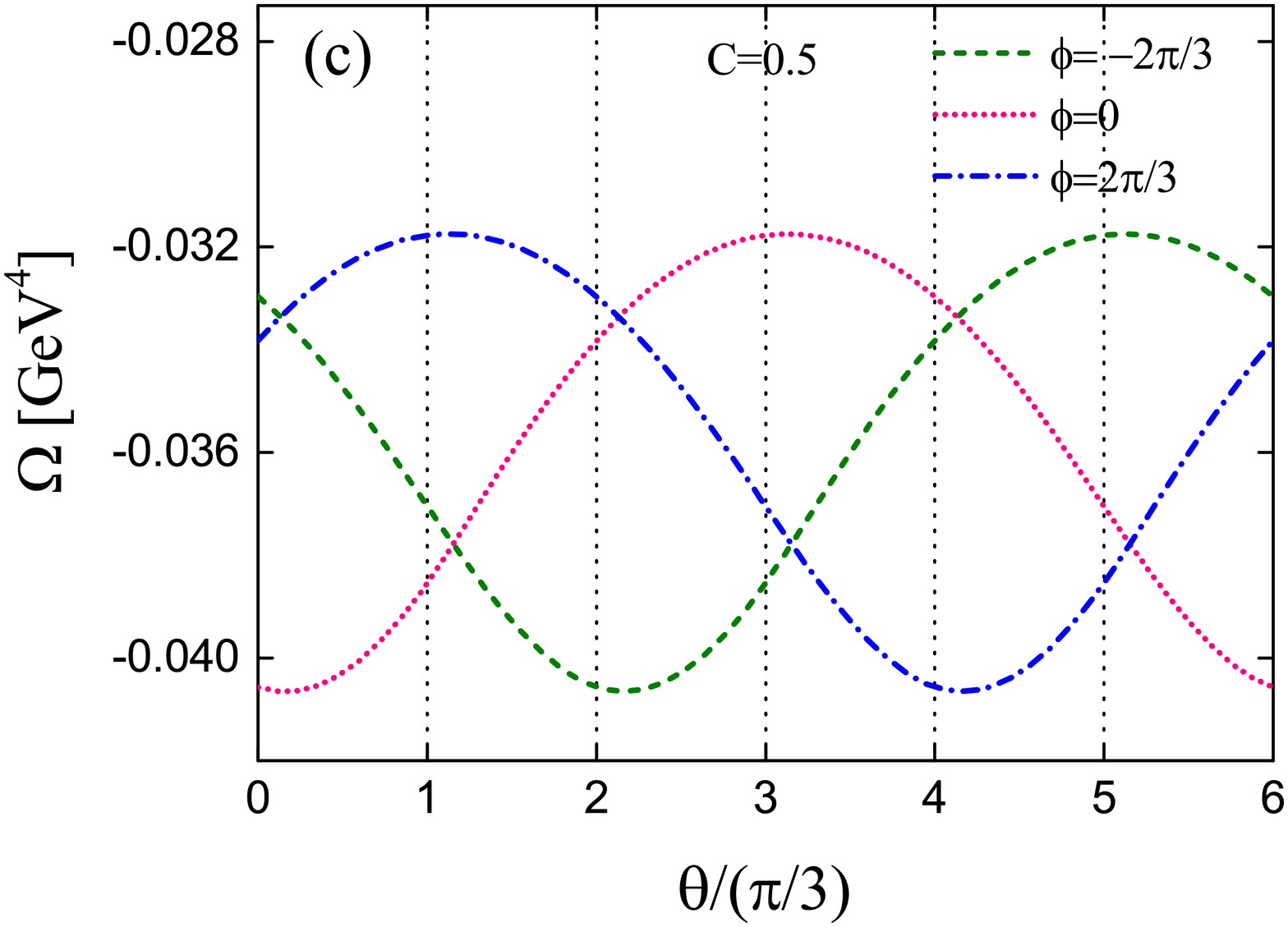}
 \includegraphics[width=0.9\columnwidth]{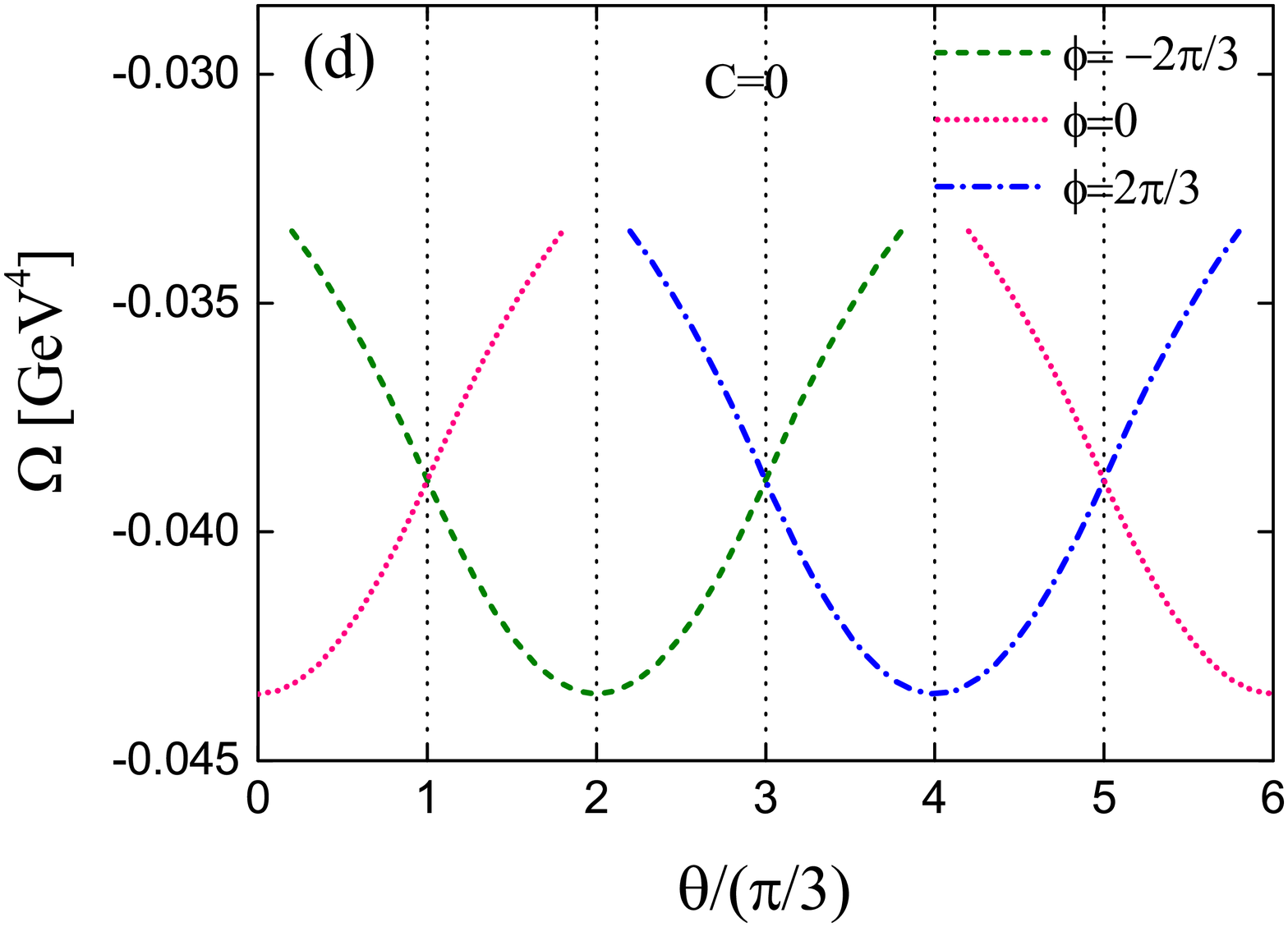}
 \caption{Thermodynamic potentials of the $\mathbb{Z}_3$ sectors for different $C$ at $T=250$ MeV in the case of $N_f=2+1$
with $m_u=m_d=5.5$ MeV and $m_s=140.7$ MeV. The RW transition point moves from $2k\pi/3$  to $(2k-1)\pi/3$
when $C$ changes from 1 to 0.}
\label{fig:11}
\end{figure*}

The upper panel of Fig.~\ref{fig:10} presents the $\theta$-dependences of $\Omega$ at $T=250$ MeV
for fixed $m_{u(d)}=140.7$ MeV and varied $m_s$ ($m_s<m_{u(d)}$). As anticipated, the cusps of $\Omega$
become sharper with the decrease of $m_s$. The lower panel shows the $\theta$-$T$ phase diagrams under
the same conditions. Similar to Fig.~\ref{fig:7}, the deconfinement transitions are all first-order,
which suggests the center symmetry breaking due to mass deference is weak. We see that $T_{RW}$
increases with the decrease of $m_s$. This means the higher the degree of center symmetry breaking,
the higher the $T_{RW}$. This point is distinct with that shown in Figs.~\ref{fig:3} and ~\ref{fig:7}
and whether it is a model artifact is unclear.

\subsection{Center symmetry breaking pattern (iv): $N_f=2+1$ with varied $C\neq {1}$}

Figure.\ref{fig:11} shows the $\theta$-dependences of $\Omega_{\phi}$ at $T=250$ MeV for different
$C$ ($C=0.9, 0.7, 0.5, 0$) with the physical quark masses. In these cases, the original center symmetry
of $\mathbb{Z}_3$-QCD is explicitly broken by both the mass difference and $C\neq{1}$.

Note that for $N_f=2+1$, the relation $\Omega(\theta)=\Omega(-\theta)$, which is true for $C=0$ or $1$,
does not hold when $C\in(0,1)$. Correspondingly, the angle $\theta_{rw}$ for $0<C<1$ is located between $(2k-1)\pi/3$
and $2k\pi/3$, which moves towards $(2k-1)\pi/3$ ($2k\pi/3$) when $C\rightarrow{0}$ ($C\rightarrow{1}$),
as demonstrated in Fig.~\ref{fig:11}.
This figure clearly shows that the tip of the cusp becomes sharper with the decrease of $C$ and thus the
standard RW transition (Fig.~\ref{fig:11}(d) represents the standard RW transition) is the strongest.

We don't plot the $\theta-T$ phase diagrams for this pattern with the physical quark masses. The traditional
RW endpoint in PNJL with physical quark masses is a triple point~\cite{Sugano}. So we can expect that the phase
diagrams for different $C's$ are similar to Fig.~\ref{fig:3} and the RW endpoints are triple ones, except
$\theta_{RW}\neq{k\pi/3}$.

\begin{figure}[htbp]
\includegraphics[width=0.9\columnwidth, clip]{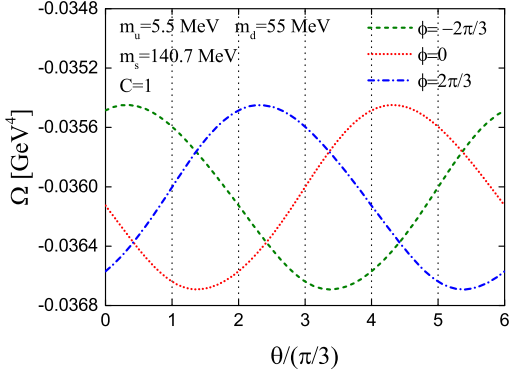}
\caption{ Thermodynamic potentials of the $\mathbb{Z}_3$ sectors at $T=250$ MeV for $C=1$
in the case of $N_f=1+1+1$ with $m_u=5.5$ MeV, $m_d=55$ MeV and $m_s=140.7$ MeV.}
\label{fig:12}
\end{figure}

\subsection{Center symmetry breaking pattern (v): $N_f=1+1+1$ with $C={1}$}

The $\theta$-dependences of $\Omega_{\phi}$ at $T=250$ MeV in pattern (v) are shown in Fig.~\ref{fig:12},
where $m_u=5.5$ MeV, $m_d=55$ MeV and $m_s=140.7$ MeV are adopted. Similar to patterns (ii)-(iii), the
original center symmetry of $\mathbb{Z}_3$-QCD is explicitly broken due to mass non-degeneracy.

Note that $\Omega(\theta)\neq\Omega(-\theta)$ in this pattern since different flavors have different masses.
As a result, the RW transitions don't occur at $\theta=k\pi/3$. Fig.~\ref{fig:12} shows that $\theta_{rw}$
is in between $2k\pi/3$ and $(2k+1)\pi/3$, which is different from Fig.~\ref{fig:11}. Fixing $m_u$ and $m_s$
and keeping $m_u<m_d<m_s$, we verify that the RW point moves towards $2k\pi/3$ ($(2k+1)\pi/3$) when
$m_d\rightarrow{m_u}$ ($m_s$). This is easily understood since the condition $m_u=m_d<m_s$ ($m_u<m_d=m_s$)
with $C=1$ corresponds to pattern (ii) ((iii)).

In pattern (v), how the RW transition depends on quark masses is complicated and the PNJL model is only
suited to study the system with light quarks. Fig.~\ref{fig:12} shows that the strength of the RW transition
is similar to cases of pattern (ii) with $m_{u(d)}=5.5$ MeV and $m_s=140.7$ MeV and pattern (iii) with
$m_{u(d)}=140.7$ MeV and $m_s=5.5$ MeV. This may suggest that in light flavor cases the RW transition due
to mass-nondegeneracy is quite weaker than the traditional RW transition, and thus the center symmetry
breaking is not so severely.

\section{Discussion and conclusion}
\label{Sec_4}

In this paper, we use the three flavor PNJL as a $\mathbb{Z}_3$-QCD model to investigate the nature 
of RW and deconfinement transitions by breaking the center symmetry in different patterns. The FTBCs 
are adopted, which correspond to the flavor-dependent imaginary chemical potentials 
$(\mu_{u},\mu_{d},\mu_{s})/iT =(\theta-2C\pi/3, \theta, \theta+2C\pi/3)$. The center symmetry of 
$\mathbb{Z}_3$-QCD is explicitly broken when three flavors are mass-nondegenerate or/and $C\neq1$.

We first demonstrate that the thermodynamic potential $\Omega(\theta)$ for $N_f=3$ and $C=1$ peaks at
$\theta=(2k+1)\pi/3$ and $2k\pi/3$ ($k\in\mathbb{Z}$) in low and high temperatures, respectively. Namely, 
the shift of the peak position of $\Omega(\theta)$ from $\theta=(2k+1)\pi/3$ to $\theta=2k\pi/3$ with $T$
just corresponds to the true first order deconfinement transition. There is no the RW transition in this 
case because of the exact center symmetry.

For $N_f=3$ with $C\neq1$, the RW transitions occur at $\theta=(2k+1)\pi/3$ when $T>T_{RW}$. The transition 
strength becomes stronger when $C$ decreasing from one and the strongest corresponds to the traditional RW 
transition with $C=0$. We verify that the RW endpoint is always a triple point in the light flavor case with
$m_u=m_d=m_s=5.5$ MeV. The corresponding first-order deconfinement transition line in the $\theta-T$ plane
becomes shorter when $C$ approaching zero. For $C$ near zero, the first-order deconfinement transition only
appears at a very small region around the RW endpoint.

For $N_f=2+1$ with $C=1$, the RW transitions appear at $\theta=2k\pi/3$ rather than $\theta=(2k+1)\pi/3$.
We argue that the angle $\theta_{RW}$ in this case is mainly determined by the two mass-degenerate light
flavors, which is supported by the previous study for the two flavor system at nonzero imaginal baryon 
and isospin chemical potentials~\cite{Sakai:2009vb}. The only heavier flavor affects the $T_{RW}$ and RW 
strength directly. For $m_u=m_d=5.5$ MeV, it is found that the tips of RW cusps become sharper with 
$m_s$ ($m_s>m_u$); moreover, the RW endpoints are always triple points and only the first-order deconfinement 
transition appears in the $\theta-T$ plane.

In contrast, the RW transitions for $N_f=1+2$ and $C=1$ still appear at $\theta=(2k+1)\pi/3$. This is because
the $\theta_{RW}$ in this pattern is determined by the only light flavor rather than the two degenerate heavier
ones, which is consistent with the argument mentioned above. Similarly, the flavor mass mismatch impacts on 
the $T_{RW}$ and RW strength significantly. For $m_u=m_d=140.7$ MeV and $m_s<m_u$, it is found that with the 
decrease of $m_s$, the RW transition gets stronger but the $T_{RW}$ becomes higher. The latter is unusual in 
comparison with the aforementioned two cases and the reason is unclear. Similar to the pattern of $N_f=2+1$ 
and $C=1$, the deconfinement transition is always first-order.

In above three patterns, the relation $\Omega(\theta)=\Omega(-\theta)$ holds and the $\theta_{RW}$'s are
integral multiples of $\pi/3$. In general, $\Omega(\theta)$ is not $\theta$-even and $\theta_{RW}$ can be
other values. For $N_f=2+1$ but $C\in(0,1)$, the $\theta_{RW}$ is located in $((2k-1)\pi/3,2k\pi/3)$, which moves
to $2k\pi/3$ ($(2k-1)\pi/3$) when $C$ approaching one (zero). In this pattern, the RW strength is more 
sensitive to the deviation of $C$ from one rather than the mass difference. In contrast, for $N_f=1+1+1$ 
with $C=1$, the $\theta_{RW}$ is located in $(2k\pi/3,(2k+1)\pi/3)$, which moves towards $2k\pi/3$
($(2k+1)\pi/3$) when $N_f=1+1+1 \rightarrow N_f=2+1$ ($N_f=1+2$).

Our calculation suggests that the deconfinement transition always keeps first-order for $C=1$ with or without
the mass degeneracy in PNJL. This indicates that the center symmetry breaking caused purely by mass difference
is too weak to lead to deconfinment crossover if the common difference of $\mu_f/{iT}$ series is $2\pi/3$ in this
model. In contrast, when $C$ deviates from one and below some critical value $C_c(\theta)$, the crossover for
deconfinement occurs at $\theta$ far from $\theta_{RW}$, which implies the strong center symmetry breaking. The
first-order deconfinement transition line in the $\theta-T$ plane shrinks with the decrease of $C$ up to zero.
Thus the strongest deconfinement transition happens at $\theta_{RW}$ and $C=0$.

The study gives predictions of how the RW and deconfinement transitions depend on the degree and manner of the 
center symmetry breaking related to $\mathbb{Z}_3$-QCD. These results may be illuminating to understand the 
relationship between $\mathbb{Z}_3$ symmetry, RW transition and deconfinement transition at finite imaginary
chemical potential and temperature region where the LQCD simulations are available. The conclusions obtained here 
are mainly based on the effective model analysis, which should be checked by other methods. Moreover, the quark 
masses can not be large enough in our calculation and it is unclear how the RW transition depends on the center 
symmetry breaking for the heavy quark system. The further study is necessary by employing the LQCD simulations 
or the perturbative strong coupling QCD by taking into FTBDc.

\vspace{5pt}
\noindent{\textbf{\large{Acknowledgements}}}\\
This work was supported by the NSFC ( No. 11875127, No.11275069 ).


\end{document}